%% file: main.tex
\documentclass[DIV=15, bibliography=totoc, twocolumn]{scrartcl}  

\usepackage[a4paper,margin=1.5cm,bottom=2.5cm]{geometry}
\usepackage{amsmath,amssymb}
\usepackage{algorithmicx}
\usepackage{algpseudocode} 
\usepackage{algorithm}

\usepackage{multicol}
\usepackage{nomencl}
\usepackage{ulem}

\makenomenclature

\renewcommand{\nompreamble}{\begin{multicols}{2}}
\renewcommand{\nompostamble}{\end{multicols}}
\usepackage{float}
\input{configuration} 
\input{00_frontpage}

\crefname{paragraph}{paragraph}{paragraphs}
\Crefname{paragraph}{Paragraph}{Paragraphs}

\begin{document}
\twocolumn[
  \begin{@twocolumnfalse}
  
    \vspace{-1.3cm} 
    \normalem \maketitle  
    \normalfont\fontsize{11}{13}\selectfont
    
        
    \vspace{-1.5cm} \hrule 
    \section*{Abstract}
Fatigue simulation requires accurate modeling of unloading and reloading. However, classical ductile damage models treat deformations after complete failure as irrecoverable --- which leads to unphysical behavior during unloading. This unphysical behavior stems from the continued accumulation of plastic strains after failure, resulting in an incorrect stress state at crack closure. As a remedy, we introduce a \textit{discontinuity strain} in the additive elasto-plastic strain decomposition, which absorbs the excess strain after failure. This allows representing pre- and post-cracking regimes in a fully continuous setting, wherein the transition from the elasto-plastic response to cracking can be triggered at any arbitrary stage in a completely smooth manner. Moreover, the presented methodology does not exhibit the spurious energy release observed in hybrid approaches. In addition, our approach guarantees mesh-independent results by relying on a characteristic length scale --- based on the discretization's resolution. We name this new methodology the \textit{discontinuous strain method}. The proposed approach requires only minor modifications of conventional plastic-damage routines. To convey the method in a didactic manner, the algorithmic modifications are first discussed for one- and subsequently for two-/three-dimensional implementations. Using a simple ductile constitutive model, the discontinuous strain method is validated against established two-dimensional benchmarks. The method is, however, independent of the employed constitutive model. Elastic, plastic, and damage models may thus be chosen arbitrarily. Furthermore, computational efforts associated with the method are minimal, rendering it advantageous for accurately representing low-cycle fatigue but potentially also for other scenarios requiring a discontinuity representation within a plastic-damage framework. An open-source implementation is provided to make the proposed method accessible.

    \vspace{0.25cm}
    \noindent\textit{Keywords:} 
    Implicit discontinuity; Strain decomposition; Unilateral effects; Crack closure; Discontinuity strain 
    \vspace{-0.4cm}
    
    \nomenclature{$\boldsymbol{\varepsilon}$, ${\varepsilon}$}{total strain}
    \nomenclature{$\boldsymbol{\varepsilon}^e$, ${\varepsilon}^e$}{elastic strain}
    \nomenclature{$\boldsymbol{\varepsilon}^p$, ${\varepsilon}^p$}{plastic strain}
    \nomenclature{$\boldsymbol{\varepsilon}^d$, ${\varepsilon}^d$}{discontinuity strain}
    \nomenclature{$\boldsymbol{\sigma}$, ${\sigma}$}{total stress}
    \nomenclature{$\boldsymbol{\sigma}_t$, ${\sigma}_t$}{tensile stress}
    \nomenclature{$\boldsymbol{\sigma}_c$, ${\sigma}_c$}{compressive stress}
    \nomenclature{$\hat\sigma_i$}{$i$-th principal stress}
    \nomenclature{$\sigma_y$}{yield stress stress}
    \nomenclature{$\boldsymbol{s}$}{deviatoric stress}
    \nomenclature{${I}_1$, ${J}_2$}{stress invariants}
    \nomenclature{$p$}{hydrostatic stress}
    \nomenclature{$q$}{von Mises stress}
    \nomenclature{$d$}{damage index}
    \nomenclature{$\mathcal{G}_f$}{fracture energy}
    \nomenclature{${\boldsymbol{D}}$}{constitutive matrix}
    \nomenclature{$E$}{Young's modulus}
    \nomenclature{$K$}{bulk modulus}
    \nomenclature{$G$}{shear modulus}
    \nomenclature{$\nu$}{Poisson's ratio}
    \nomenclature{$\beta$}{dilation constant}
    \nomenclature{$k$}{damage internal variable}
    \nomenclature{$d_c$}{critical damage}
    \nomenclature{$k_c$}{critical damage internal variable}
    \nomenclature{$\dot{\gamma}$}{plastic multiplier}
    \nomenclature{$\ell$}{length scale}
    \nomenclature{$\mathcal{P}$}{projection tensor}
    \nomenclature{$\mathcal{I}$}{fourth-order identity tensor}
    \nomenclature{$\boldsymbol{I}$}{second-order identity tensor}
    \nomenclature{$\alpha$}{damage growth constant}
    \nomenclature{$g(\boldsymbol{\sigma})$}{Drucker--Prager flow potential}    
    \nomenclature{$f(\boldsymbol{\sigma})$}{Rankine yield function}
    
    \printnomenclature
    \vspace{10pt}
    \end{@twocolumnfalse}
]

\section{Introduction}\label{sec:introduction}
Computational failure analysis is ideally approached by a combination of continuous and discontinuous methods. The reason lies in the fact that neither of them is adequate to describe the entire failure process on its own. Continuous methods suit the early stages of failure during which the material degradation is of a diffuse type. No predefined imperfection or crack is needed, no remeshing procedure is required, and no discontinuous enrichment function is involved. However, classical continuous methods fail to provide an exact representation of cracks since discontinuities cannot be treated discretely --- instead they are modeled implicitly by smearing their effect over a small region which mimics the behavior by a locally adjusted constitutive law. Discontinuous methods, on the other hand, excel in simulating sharp cracks emerging due to excessive material degradation. Instead of smearing strong discontinuities over a finite width of medium they provide the precise trajectories of discontinuities and the associated displacement jump. Despite their desirable properties, discontinuous methods face difficulties in dealing with the diffuse material degradation occurring over the fracture process zone. Hence, a valid compromise is a hybrid method that treats the early stage of fracture, i.e. the degrading of material, in the framework of continuum mechanics and represents fully degraded regions through discontinuities. Although such a continuous-discontinuous method is well-equipped to deal with the two extreme ends of the failure process, a smooth transition from the continuous to the discontinuous model is additionally needed to prevent spurious energy generation or dissipation. Given the aforementioned complexities, computational failure analysis is a challenging task and still subject to ongoing research.

\subsection{Continuous Models}

The failure process is triggered by the nucleation of micro-cracks, which causes the load-bearing capacity as well as the stiffness of the material to degrade. Continuum damage mechanics is tailored to describe such phenomena in a smeared sense so that only the density of those micro-cracks is considered. Early constitutive models in the framework of continuum damage mechanics rely on a scalar damage index that characterizes that density~\cite{lemaitre1985continuous, krajcinovic1989damage}. Due to different dissipative phenomena, the material degradation can be described by means of multiple variable damage indices~\cite{ladeveze1983anisotropic}. These damage indices were replaced by damage tensors in more sophisticated anisotropic models so that damage on different planes can be described with different densities~\cite{leckie1981tensorial, chow1987anisotropic}. Crack closure effects were also introduced to the theory, enabling it to reproduce the unilateral behavior upon the closing and re-opening of micro-cracks~\cite{ladeveze1984damage, desmorat2000quasi}. As the softening response induced by the growth of imperfections leads to local material instability, the boundary value problem must be regularized. Otherwise, the uniqueness of solutions cannot be guaranteed. The most straightforward remedy to circumvent the complexities arising from regularization methods is to adjust the constitutive behavior of each integration point in accordance with the spatial discretization such that the total energy dissipation in the medium remains constant~\cite{bavzant1983crack}. However, if one is obliged to sacrifice simplicity for the sake of accuracy, one must opt for a regularization technique such as non-local or gradient-enhanced methods. Non-local formulations introduce redistribution effects in an integral sense so that, regardless of the numerical discretization, the fracture process zone spans over a prescribed width of the medium~\cite{bazant2002nonlocal, amor2009regularized}. Conversely, gradient-enhanced models use Helmholtz-like differential equations to incorporate these diffusion effects~\cite{peerlings1996gradient, seupel2018efficient}. By considering an additional field variable alongside the displacement field, gradient-enhanced models are considered as mixed formulations opening a variety of possibilities for modeling the failure process~\cite{dias2016288}. This additional field either occupies the whole domain in classical mixed formulations~\cite{voyiadjis2019strain, brepols2020102635}, or is restricted to specific subdomains in the so-called \textit{strain-injection} technique~\cite{oliver2014289, oliver2015384, dias2016288, lloberasvalls2016499, dias2018, dias2018b}. In analogy to the gradient-enhanced damage formulation, phase-field models introduce the redistribution effects by means of a set of partial differential equations, which must be solved alongside the original boundary value problem of the system~\cite{schreiber2021phase, yan2022efficient, Kuhn2022, phansalkar2022spatially}. Many different applications have been developed based on phase-field models, such as fracture in brittle materials~\cite{kuhn2010continuum, hug_3d_2020}, ductile failure~\cite{miehe2016phase}, finite strain analysis~\cite{bilgen2021phase, Bilgen2022}, dynamic problems~\cite{schluter2014phase, weinberg2022dynamic, partmann2023continuum}, structural elements~\cite{gebuhr2022damage, pise_phenomenological_2023}, composite materials~\cite{storm2021comparative}, and fatigue analysis~\cite{alessi2018phenomenological, schroder2022phase}, to name a few.

\subsection{Discontinuous Models}

Nonlinear fracture mechanics is defined on an entirely different basis. The fracture process zone is replaced by a fictitious crack that extends along the existing crack, and its nonlinear response is incorporated by applying surface tractions on the opposing faces of the extended crack~\cite{moes2002extended}. The resulting method, known as the cohesive crack method, intrinsically regularizes the boundary value problem by replacing the softening region with a set of measure zero~\cite{rabczuk2013computational}. Hence, it comes at the cost of defining strong discontinuities by which the displacement jump across the cohesive zone can be mimicked. Needless to say, the discretized boundary value problem must be updated in accordance with the configuration of propagating cracks. This update can either be performed by aligning the spatial discretization along the crack or by means of introducing enriched basis functions through, e.g., the extended finite element method (XFEM). The cohesive crack method has been used in many applications, namely for quasi-brittle materials~\cite{tijssens2000numerical}, elastic-plastic cracking~\cite{tvergaard1996effect}, dynamic crack propagation~\cite{camacho1996computational}, delamination analysis~\cite{turon2007engineering}, hybrid components~\cite{toller2019bulk}, and membrane analysis~\cite{toller2020applying}. XFEM, which relies on enriched basis functions, circumvents the remeshing problem yet still manipulates the global algebraic system. Applications of the method include but are not limited to, crack growth modeling in concrete~\cite{unger2007modelling}, interface failure analysis~\cite{kastner2013xfem, kastner2016xfem}, crack propagation in fiber-reinforced composites~\cite{pike2015xfem}, and biomechanical problems~\cite{idkaidek2018fracture}.

\subsection{Hybrid Models}

\begin{figure*}
    \centering
    \includegraphics[scale=1.0]{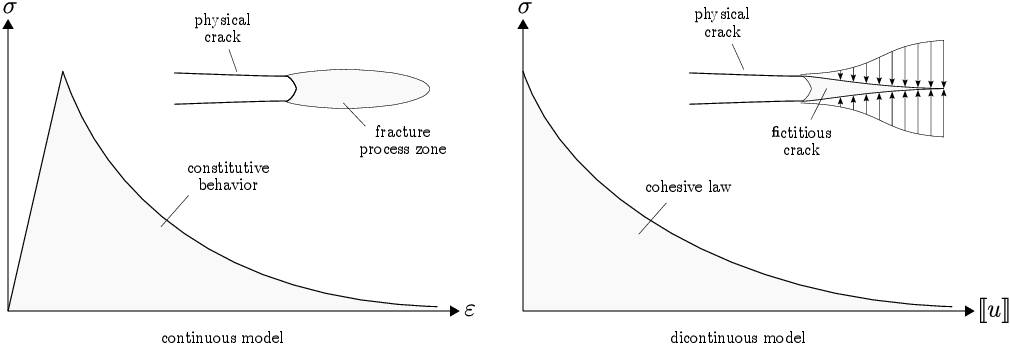}
    \caption{Analogy between constitutive behavior and cohesive law in continuous and discontinuous models.}
    \label{fig:transition}
\end{figure*}

Many authors have opted for a hybrid approach that combines the versatility of continuous models in the pre-cracking phase and the robustness of discontinuous ones in the post-cracking regime. The continuous-discontinuous gradient-enhanced damage model~\cite{simone2003continuous}, thick level set technique~\cite{moes2011level}, combined XFEM–damage mechanics model~\cite{sun20213d}, and thin layer approach~\cite{puccia2023finite} are some examples of such methods. Besides facing the complexities of both techniques, a transition phase must be defined such that no spurious energy transfer occurs. Some authors circumvent this by minimizing the energy release by injecting the strong discontinuity when the load-bearing capacity of the material is almost zero~\cite{seabra2013damage, sarkar2021simplified}. By contrast, others trigger the transition at an intermediate stage at the cost of solving the energy transformation issue~\cite{cuvilliez2012finite, roth2015combined}. However, a fully smooth transition is almost impossible as some portion of energy must dissipate in the continuum settings, and the remaining part must be transferred to the discontinuous model. \Cref{fig:transition} shows the analogy between the constitutive behavior in continuous models and a cohesive law in discontinuous approaches. It is worth mentioning that the area under the curve in each setting represents the fracture energy $\mathcal{G}_f$, which is the total energy that is required to create a unit area of a crack. Once the transition stage is reached, a fictitious crack in the form of a strong discontinuity must be injected such that no imbalance between internal and external forces occurs. Moreover, the sum of dissipated energy in the continuous setting and the remaining energy in the discontinuous model must be equal to the fracture energy. Note that the energy dissipation in a continuous model occurs over the width of the fracture process zone. At the same time, the dissipative mechanism is concentrated on two opposing points on the fictitious crack faces in a discontinuous model. Therefore, it is questionable whether the energy dissipated across the width of the fracture process zone must be used to define the remaining part or if a point-to-point transition would be valid. In addition, the constitutive behavior in a continuous model provides the stress-strain relation, while the constitutive law in discontinuous models defines the traction-separation law. Numerically speaking, since the strain field is obtained by differentiating the displacement field, the strain variation across the fracture process zone and the variation of the displacement jump along the fictitious crack are of different orders of smoothness. Hence, injecting a strong discontinuity along the previously damaged region of the model, i.e., in the fracture process zone, is accompanied by major inconsistencies that cause spurious energy imbalances.

\subsection{Discontinuous Strain Method}

\begin{figure*}
    \centering
    \includegraphics[scale=1.0]{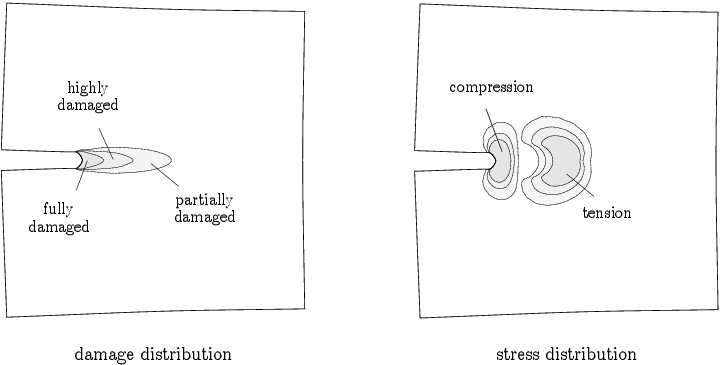}
    \caption{Damage distribution, indicated in terms of the damage index $d$, and stress distribution in an edge-notched rectangular plate under a relaxed state after experiencing the opening mode of fracture.}
    \label{fig:artificial}
\end{figure*}

The method we are presenting fully relies on continuous approaches in the sense that no strong discontinuity or additional degrees of freedom are needed. Nevertheless, it benefits from the robustness of discontinuous methods. To this end, we use an additional strain field that mimics strong discontinuities. This augmentation is essential in ductile damage formulations within which irreversible straining is considered alongside the reversible elastic one. To elaborate on this necessity, schematical distributions of damage and stress in an edge-notched rectangular plate under a relaxed state --- after experiencing the opening mode of fracture --- are presented in \Cref{fig:artificial}. The damage index asymptotically reaches unity across the fully damaged region. This area represents the region where material has been detached due to excessive damage growth yet is kept connected in a continuous setting. As a result, this region experiences unrealistic plastic deformations. Hence, it can still withstand pressure if the unilateral effects arising from a load reversal become active~\cite{daneshyar_ductile_2023}. 

By injecting the discontinuity strain field at an intermediate state, further straining is accompanied by the growth of the discontinuity strain so that the excessive unrealistic permanent straining is prevented. In contrast to the plastic strain, the discontinuity strain is reversible. Hence, no artificial straining can occur after removing the external load. In addition, once the faces of the implicit crack meet, the unilateral effects become active, and the crack closing and reopening process is reproduced. Since both pre- and post-cracking regimes are defined in a continuous setting, the transition stage at which the discontinuity strain is mobilized can be chosen arbitrarily without introducing any force imbalance or energy release. By employing the concept of energy equivalence, the constitutive behavior of the material is adjusted for each integration point such that the global response remains objective to the spatial discretization of the domain. 
Accordingly, the remainder of this work is organized as follows. In \Cref{sec:theory}, the nonlinear material model incorporating the additive strain decomposition extended by the discontinuity strain field, the plasticity model, the damage evolution, and the energy equivalence concept are presented. \Cref{sec:implementation} is dedicated to the implementation of the presented model in one- and multi-dimensional spaces. Numerical validation, mesh sensitivity, and algorithmic overhead of the model are discussed in \Cref{sec:results}. Finally, some conclusions are drawn in \Cref{sec:conclusion}.
 
\section{Theory}\label{sec:theory}

\subsection{Additive Strain Decomposition}\label{ssec:strainDecomposition}
\begin{figure*} [htb]
    \centering
    \includegraphics[scale=1.0]{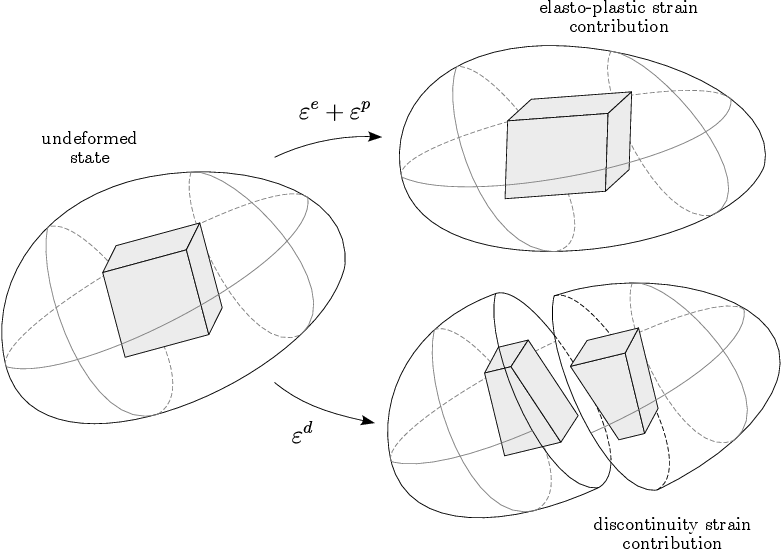}
    \caption{Schematics of additive strain decomposition augmented by the discontinuity strain field.}
    \label{fig:infinitesimal}
\end{figure*}

The presented model relies on augmenting the conventional additive strain decomposition in the infinitesimal strain theory with an additional field, called the discontinuity strain, so that the unilateral effects arising from the interaction of the opposing faces of strong discontinuities can be captured within a fully continuous setting. Hence, we define the following strain decomposition
\begin{equation}
    \boldsymbol{\varepsilon} = \boldsymbol{\varepsilon}^e+\boldsymbol{\varepsilon}^p+\boldsymbol{\varepsilon}^d\label{eq:strainDecomposition},
\end{equation}
where $\boldsymbol{\varepsilon}$, $\boldsymbol{\varepsilon}^e$, $\boldsymbol{\varepsilon}^p$, and $\boldsymbol{\varepsilon}^d$ are the total, elastic, plastic, and discontinuity strain tensors respectively. This decomposition is schematically depicted in \Cref{fig:infinitesimal}. The discontinuity strain field is responsible for mimicking the strain jump across the implicit discontinuous interface.

In the framework of continuum damage mechanics, two different definitions exist for the total stress tensor: (i) the effective stress tensor and (ii) the total stress tensor. The elasto-plastic constitutive behavior is defined in a so-called \textit{effective} configuration in which the material is assumed to be undamaged, and the degradation effects are incorporated by mapping the stress state, i.e., the effective stress $\tilde{\boldsymbol{\sigma}}$ to its counterpart in the damaged configuration, which we in this paper refer to as the total stress $\boldsymbol{\sigma}$. Doing so, the effective stress $\tilde{\boldsymbol{\sigma}}$ is given by means of the linear elastic constitutive law as follows
\begin{align}
    \begin{split}
    \tilde{\boldsymbol{\sigma}}&=\tilde{\boldsymbol{D}}:\boldsymbol{\varepsilon}^e\\
    &=\tilde{\boldsymbol{D}}:(\boldsymbol{\varepsilon}-\boldsymbol{\varepsilon}^p-\boldsymbol{\varepsilon}^d)\label{eq:elasticConstitutiveLaw},
    \end{split}
\end{align}
where $\tilde{\boldsymbol{D}}$ is the constitutive matrix of intact material. Unless otherwise noted, the tilde accent over a quantity refers to the undamaged configuration.

\subsection{Tension-Compression Stress Split}\label{ssec:stressSplit}
Material degradation arising from tensile micro-cracking reduces the load-bearing capacity as well as the stiffness of the material. However, those reductions vanish if a compressive load is applied. This phenomenon can be attributed to the crack closure effects. In this regard, one of the main aspects in constitutive modeling of material degradation is reproducing the unilateral effects arising from the crack closing and reopening. To this end, we split the effective stress tensor into
\begin{equation}
    \tilde{\boldsymbol{\sigma}}=\tilde{\boldsymbol{\sigma}}_t + \tilde{\boldsymbol{\sigma}}_c\label{eq:tensionCompressionSplit},
\end{equation}
where $\tilde{\boldsymbol{\sigma}}_t$ and $\tilde{\boldsymbol{\sigma}}_c$ are the tensile and compressive parts, respectively. The latter can be given by
\begin{equation}
    \tilde{\boldsymbol{\sigma}}_t = \mathcal{P}:\tilde{\boldsymbol{\sigma}}\label{eq:tensionStress},
\end{equation}
where $\mathcal{P}$ is the projection tensor
\begin{equation}
    \mathcal{P} = \sum_{i=1}^3{H(\tilde\sigma_i)(\boldsymbol{n}_i\otimes\boldsymbol{n}_i\otimes\boldsymbol{n}_i\otimes\boldsymbol{n}_i)}, \\\label{eq:projectionTensor}
\end{equation}
wherein
\begin{equation}
    H(x) = \begin{cases} 1, \qquad x>0\\ 0, \qquad x\leq 0 \end{cases}
\end{equation}
is the Heaviside function, $\tilde\sigma_i$ is the $i$-th effective principal stress, and $\boldsymbol{n}_i$ is its corresponding eigenvector~\cite{ortiz1985constitutive}. Utilizing \Cref{eq:tensionCompressionSplit}, the compressive part reads
\begin{equation}
    \tilde{\boldsymbol{\sigma}}_c = \tilde{\boldsymbol{\sigma}} - \mathcal{P}:\tilde{\boldsymbol{\sigma}},
\end{equation}
or, equivalently,
\begin{equation}
    \tilde{\boldsymbol{\sigma}}_c = (\mathcal{I} - \mathcal{P}):\tilde{\boldsymbol{\sigma}},
\end{equation}
where $\mathcal{I}$ is the fourth-order identity tensor.

With the tensile and compressive parts at hand, the stress tensor in the damage configuration is defined as
\begin{equation}
    \boldsymbol{\sigma}=(1-d)\tilde{\boldsymbol{\sigma}}_t + \tilde{\boldsymbol{\sigma}}_c\label{eq:trueStress},
\end{equation}
where $d$ is called the damage index. Accordingly, the material degradation only affects the tensile part of the stress, while the compressive part remains intact. Note that this definition enables the stiffness recovery during the transition from tensile to compressive loading. Hence, it introduces the crack closure effects to the model.

\subsection{Plasticity Model}\label{ssec:plasticityModel}
Since only the tensile cracking is of interest, we use the Rankine maximum stress criterion to define the yield function, so that
\begin{equation}
    f(\tilde{\boldsymbol{\sigma}}) = \hat{\tilde{\sigma}}_\text{max} - \sigma_y\label{eq:yieldSurface},
\end{equation}
where $\hat{\tilde{\sigma}}_\text{max}$ is the maximum principal stress in the effective configuration, and $\sigma_y$ is the yield stress.

According to the plasticity theory, the stress state must remain inside or on the yield surface. This implies that the yield function $f$ can be either negative or zero. Violating this admissibility mobilizes the plastic flow such that the stress state returns to the yield surface. Note that although the Rankine maximum stress criterion could be an acceptable choice for frictional materials such as rocks and concretes, it leads to unacceptable inelastic dilatation if it is used in an associative plasticity model. A common choice for those materials is to use a Drucker--Prager type potential function so that the proper dilatancy that is observed in their inelastic response can be achieved~\cite{lee1998plastic, wu2006energy}. Thus, the plastic strain increment $\dot{\boldsymbol{\varepsilon}}^p$ is defined by means of the non-associative flow rule
\begin{equation}
    \dot{\boldsymbol{\varepsilon}}^p=\dot{\gamma}\partial_{\tilde{\boldsymbol{\sigma}}} g \label{eq:associativeFlowRule},
\end{equation}
where $\dot{\gamma}$ is the plastic multiplier and
\begin{equation}
    g(\tilde{\boldsymbol{\sigma}}) = 3\beta \tilde{p} + \tilde{q} \label{eq:plasticPotential}
\end{equation}
is the Drucker--Prager potential function, in which $\beta$ is the dilation constant, $\tilde{p}$ is the volumetric part of the effective stress tensor, and $\tilde{q}$ is the von Mises stress in the effective configuration. The latter two can be given by means of the effective stress invariants $\tilde{I}_1$ and $\tilde{J}_2$ as
\begin{align}
    \tilde{p} &= \tfrac{1}{3}\tilde{I}_1, \\
    \tilde{q} &= \sqrt{3\tilde{J}_2}.
\end{align}
As a result, the flow rule reads
\begin{equation}
    \dot{\boldsymbol{\varepsilon}}^p = \dot{\gamma}\Big(\beta\boldsymbol{I} + \frac{3}{2}\frac{\tilde{\boldsymbol{s}}}{\tilde{q}}\Big),
\end{equation}
wherein $\boldsymbol{I}$ is the second-order identity tensor, and $\tilde{\boldsymbol{s}}$ is the deviatoric part of the effective stress. Note, that the yield function $f(\tilde{\boldsymbol{\sigma}})$ and the plastic multiplier $\dot{\gamma}$ must obey the Karush--Kuhn--Tucker conditions
\begin{align}
    f\leq0 \label{eq:KKT1},\\
    \dot{\gamma}\geq0 \label{eq:KKT2},\\
    \dot{\gamma}f=0 \label{eq:KKT3}.
\end{align}

\subsubsection{Algorithmic Aspects}
Let us assume a material point in which the discontinuity strain $\boldsymbol{\varepsilon}^d$ has not been mobilized yet. Now, by implementing a pseudo-time-stepping scheme with the given total strain increment $\Delta\varepsilon$ and assuming a purely elastic loading state, the trial effective stress tensor is given as
\begin{equation}
    \tilde{\boldsymbol{\sigma}}_{n+1}^{trial} = \tilde{\boldsymbol{\sigma}}_{n} + \tilde{\boldsymbol{D}}:\Delta\boldsymbol{\varepsilon},
\end{equation}
where the subscripts $n$ and $n+1$ denote the previous (known) and updated (unknown) states, respectively. By applying the volumetric--deviatoric stress split, the trial stress tensor reads
\begin{equation}
    \tilde{\boldsymbol{\sigma}}_{n+1}^{trial} = \tilde{p}_{n+1}^{trial}\boldsymbol{I} + \tilde{\boldsymbol{s}}_{n+1}^{trial}.
\end{equation}
Alternatively, in accordance with Hooke's law, we can write
\begin{align}
    \tilde{p}_{n+1}^{trial} &= \tilde{p}_{n} + 3K\,\text{vol}(\Delta\boldsymbol{\varepsilon}), \\
    \tilde{\boldsymbol{s}}_{n+1}^{trial} &= \tilde{\boldsymbol{s}}_{n} + 2G\,\text{dev}(\Delta\boldsymbol{\varepsilon}),
\end{align}
where $K$ and $G$ are the bulk and shear modulus, respectively. Now, if the trial stress violates the admissibility condition 
\begin{equation}
    f(\tilde{\boldsymbol{\sigma}}_{n+1}^{trial}) \leq 0 ,
\end{equation}
the total strain increment $\Delta\boldsymbol{\varepsilon}$ mobilizes the plastic flow and the trial state must return on the yield surface. Hence, the volumetric and deviatoric parts are updated in accordance with the flow rule in a fully implicit backward Euler scheme as follows
\begin{align}
    \tilde{p}_{n+1} &= \tilde{p}_{n+1}^{trial} - 3K\Delta\gamma\beta, \\
    \tilde{\boldsymbol{s}}_{n+1} &= \tilde{\boldsymbol{s}}_{n+1}^{trial} - 3G\Delta\gamma\frac{\tilde{\boldsymbol{s}}_{n+1}}{\tilde{q}_{n+1}}.
\end{align}
We can rewrite the deviatoric part as
\begin{equation}
    \tilde{\boldsymbol{s}}^{trial}_{n+1} = \tilde{\boldsymbol{s}}_{n+1} + 3G\Delta\gamma\frac{\tilde{\boldsymbol{s}}_{n+1}}{\tilde{q}_{n+1}},
\end{equation}
yielding,
\begin{equation}
    \tilde{\boldsymbol{s}}^{trial}_{n+1} = (\tilde{q}_{n+1} + 3G\Delta\gamma)\frac{\tilde{\boldsymbol{s}}_{n+1}}{\tilde{q}_{n+1}},
\end{equation}
or, equivalently,
\begin{equation} \label{eq:colinear}
    \frac{\tilde{\boldsymbol{s}}^{trial}_{n+1}}{\tilde{q}_{n+1}^{trial}} = \frac{\tilde{\boldsymbol{s}}_{n+1}}{\tilde{q}_{n+1}},
\end{equation}
wherein
\begin{equation} \label{eq:vonUpdate}
    \tilde{q}_{n+1}^{trial} = \tilde{q}_{n+1} + 3G\Delta\gamma.
\end{equation}
A deduction that can be made from \Cref{eq:colinear} is that the tensors $\tilde{\boldsymbol{s}}^{trial}_{n+1}$ and $\tilde{\boldsymbol{s}}_{n+1}$ are co-linear~\cite{de_souza_neto_computational_2011}. Utilizing this property, the deviatoric stress can be directly updated in the space of principal stresses.

\begin{figure}
    \centering
    \includegraphics[scale=1.0]{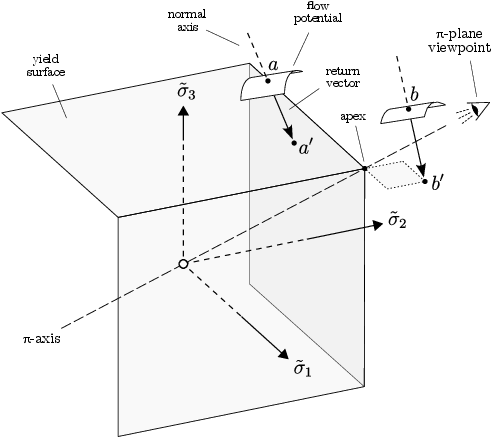}
    \caption{Schematics of return mapping in the effective Haigh--Westergaard stress space.}
    \label{fig:yield}
\end{figure}

\begin{figure}
    \centering
    \includegraphics[scale=1.0]{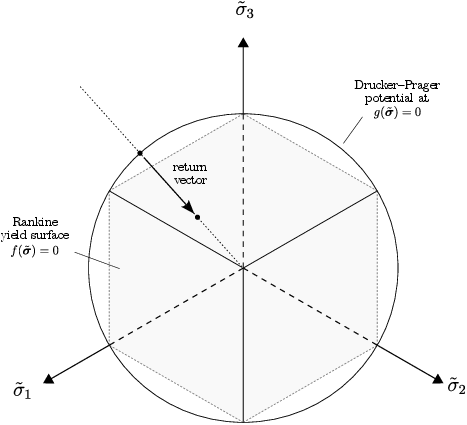}
    \caption{The $\pi$-plane representation of the yield surface and flow potential in the effective Haigh--Westergaard stress space.}
    \label{fig:normality}
\end{figure}

The schematics of the yield surface and the return-mapping procedure are shown in the effective Haigh--Westergaard stress space in \Cref{fig:yield}. The $\pi$-plane representation of the yield surface and flow potential are also presented in \Cref{fig:normality}. Note that the viewpoint in this figure is shown by an eye symbol in the three-dimensional space. According to the $\pi$-plane representation, the return-mapping procedure does not alter the arrangement of principal stresses. Hence, if yielding has occurred due to excessive stress along the $i$-th principal axis, the same principal stress remains the maximum one after the return-mapping. This significantly simplifies the procedure since only the maximum principal stress indicates whether a stress state is valid or not. Accordingly, the maximum principal stress is updated directly as follows
\begin{equation}
    \hat{\tilde{\sigma}}_{n+1} = \hat{\tilde{\sigma}}_{n+1}^{trial} - 3K\Delta\gamma\beta - 3G\Delta\gamma\frac{\hat{\tilde{\sigma}}_{n+1}^{trial}-\tilde{p}_{n+1}^{trial}}{\tilde{q}_{n+1}^{trial}},
\end{equation}
wherein the subscript ``max'' is dropped for the sake of convenience. Now, by substituting the updated maximum principal stress in the yield function and setting
\begin{equation}
    f(\tilde{\boldsymbol{\sigma}}_{n+1}) = 0 ,
\end{equation}
the plastic multiplier increment $\Delta\gamma$ is obtained \textit{explicitly} as follows
\begin{equation}
    \Delta\gamma = \frac{\hat{\tilde{\sigma}}_{n+1}^{trial} - \sigma_y}{3K\beta + 3G(\hat{\tilde{\sigma}}_{n+1}^{trial}-\tilde{p}_{n+1}^{trial})/\tilde{q}_{n+1}^{trial}}. 
\end{equation}

The issue that may arise in the stress update procedure of plasticity models with cone-shaped yield surfaces --- such as Mohr--Coulomb and Drucker--Prager --- is that the return vector may point towards an invalid state. This issue mostly arises under triaxial tension. As shown in \Cref{fig:yield}, the trial stress state located at point $a$ is properly returned to the yield surface. However, by considering the trial stress state at point $b$, the return vector points toward the invalid state $b^\prime$ which is not located on the yield surface. This issue can be simply detected by checking whether the return vector passes the $\pi$-axis or not --- possible due to the fact that the return vector is always pointing towards the $\pi$-axis (see \Cref{fig:normality}). In this case, the sign of the updated von Mises stress becomes negative, which is obviously incorrect. Hence, the upper bound of $\Delta\gamma$ can be defined using the relation from \Cref{eq:vonUpdate} as follows
\begin{equation}
    \Delta\gamma \leq \frac{\tilde{q}_{n+1}^{trial}}{3G}.
\end{equation}
If the above condition is violated, the stress state must return to the apex of the yield surface~\cite{de_souza_neto_computational_2011}. Knowing that the apex is located on the $\pi$-axis, the updated stress would be in a hydrostatic state having the stress magnitude of $\sigma_y$ in all its three principal directions. As a result, we have
\begin{align}
    \tilde{p}_{n+1}^{trial} - 3K\Delta\gamma\beta = \sigma_y,\\
    \tilde{\boldsymbol{s}}^{trial}_{n+1} - 2G\,\text{dev}(\Delta\boldsymbol{\varepsilon}^p) = \boldsymbol{0},
\end{align}
which yields
\begin{equation}
    \Delta\gamma = \frac{\tilde{p}_{n+1}^{trial} - \sigma_y}{3K\beta}
\end{equation}
and
\begin{equation}
    \text{dev}(\Delta\boldsymbol{\varepsilon}^p) = \frac{\tilde{\boldsymbol{s}}^{trial}_{n+1}}{2G}.
\end{equation}
Hence, by knowing the trial state and updated state, we can directly obtain the return vector and compute the plastic strain increment $\Delta\boldsymbol{\varepsilon}^p$ as follows
\begin{equation}
    \Delta\boldsymbol{\varepsilon}^p = \frac{\tilde{p}_{n+1}^{trial}-\sigma_y}{3K}\boldsymbol{I} + \frac{\tilde{\boldsymbol{s}}^{trial}_{n+1}}{2G}, \label{eq:solutionPlasticStrainUpdate}
\end{equation}
which avoids iterative procedures\footnote{Note that this is possible due to the choice of the non-associative flow model relying on the Rankine yield surface and the Drucker--Prager plastic potential. We are well aware that this choice might seem inferior in light of much more capable plasticity models. However, the selected plasticity model is simple and, therefore, well-suited to showcase the concepts behind the proposed discontinuous strain method.} such as those summarized in~\cite{de_souza_neto_computational_2011}.

\subsection{Damage Evolution}\label{ssec:damageEvolution}
Damage growth introduces softening effects leading to ill-posedness of the boundary value problem. As a remedy, one can opt for a variety of regularization techniques such as non-local and gradient-enhanced methods, each of which increases the computational complexity to a high extent. Alternatively, the constitutive behavior can be adjusted in accordance with the spatial discretization so that the dissipated energy is kept 
constant in a global sense. In this regard, we define a unique rate of damage growth for each integration point in accordance with its domain of influence. Hence, the area under the stress-strain curve, which represents the total energy dissipation due to complete failure, must be integrated. To facilitate the integration of the stress-strain curve, a popular choice is the exponential damage growth function
\begin{equation}
    d = 1 -\exp(-\alpha k)\label{eq:damageEvolution},
\end{equation}
where $\alpha$ is a dimensionless constant that defines the growth rate and $k$ is the damage internal variable. Based on the work of Lee and Fenves~\cite{lee1998plastic}, we define the evolution of $k$ as
\begin{equation}
    \dot{k} = w(\tilde{\boldsymbol\sigma}) \dot\varepsilon^p_\text{max}\label{eq:damageInternalPlastic},
\end{equation}
where $w$ is a weight function, defined as
\begin{equation}
        w(\tilde{\boldsymbol\sigma}) = \frac{\sum^3_{i=1}  \langle \hat{\tilde\sigma}_i\rangle}{\sum^3_{i=1} |\hat{\tilde\sigma}_i|}\label{eq:weight},
\end{equation}
and the angle brackets, known as the Macaulay brackets, denote the operator
\begin{equation}
    \langle x \rangle = \begin{cases} x, \qquad x>0\\ 0, \qquad x\leq 0. \end{cases}
\end{equation}

On the other hand, to prevent excessive plastic straining that causes the early stiffening artifact, the elastic and plastic strains are kept constant and the discontinuity strain $\boldsymbol\varepsilon^d$ is mobilized if the damage index reaches its critical threshold. At this moment, we have
\begin{equation}
    d_{c} = 1 -\exp(-\alpha k_{c}),
\end{equation}
or, equivalently,
\begin{equation}
     k_{c} = -\frac{1}{\alpha}\ln{(1-d_{c})}\label{eq:damageThreshold},
\end{equation}
where $d_{c}$ is the critical damage and $k_{c}$ is its corresponding internal variable. From this moment forward, no plastic straining occurs, and the strain jump of the hypothetical crack controls the damage evolution. Since degradation is an irreversible process, but the strain jump can increase or decrease over time, the maximum jump that is experienced during the loading history indicates the damage state. Accordingly, the damage internal variable after onset of cracking is defined as
\begin{equation}
    k = k_{c} + \max_{\tau\leq t}(\varepsilon^d_n)\label{eq:updateDamageInternalVariable},
\end{equation}
where
\begin{equation}
    \varepsilon^d_n = \boldsymbol{n}^\intercal\boldsymbol{\varepsilon}^d\boldsymbol{n} \label{eq:crackOpeningStrain}
\end{equation}
is the crack opening strain and $\boldsymbol{n}$ is the normal to the cracking plane. Since crack growth in a material point occurs on the planes with maximum tensile stress, the normal vector $\boldsymbol{n}$ is chosen to be the direction of maximum principal stress at the onset of cracking.

\subsection{Energy Equivalence}\label{ssec:energyEquivalence}

\begin{figure}
    \centering
    \includegraphics[scale=1.0]{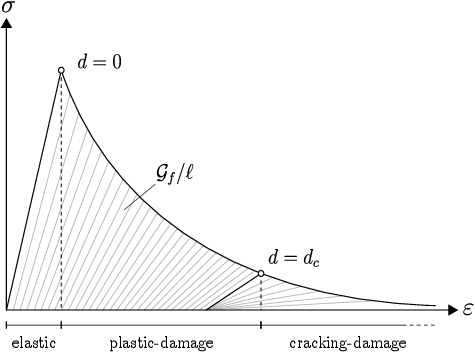}
    \caption{Stress-strain curve for a typical damage-driven softening response.}
    \label{fig:fracture_energy}
\end{figure}

A typical stress-strain curve based on exponential damage growth is given in \Cref{fig:fracture_energy}. In a continuum setting, the area under this curve is the total amount of energy. Part of the energy is dissipated (the plastic and cracking contribution), and part of it is stored (the elastic contribution) at its respective infinitesimal volume element. For materials undergoing complete failure, this energy dissipation is attributed to crack formation. According to the crack band theory of Ba\v{z}ant and Oh~\cite{bavzant1983crack}, the total energy, i.e., elastic, plastic, and cracking (see \Cref{fig:fracture_energy}) dissipated during a material point failure, can be related to the fracture energy $\mathcal{G}_f$ by considering the fact that this dissipation occurs over a finite width of the material, known as the material length scale. Since no regularization technique is used to treat the ill-posedness of the governing partial differential equations, the softened region spreads in accordance with the spatial discretization. Hence, in a finite element analysis, solutions become mesh dependent in the sense that less energy dissipation occurs if finer meshes are used, and vice versa (see \Cref{fig:case1ellDependency} in \Cref{ssec:Results2D} for the effect). By replacing the material length scale with the numerical one, the mesh-objectivity of the solutions with respect to the spatial discretization can be preserved.

Assuming a monotonic one-dimensional response under pure tension, we can write
\begin{equation} \label{eq:fracture_energy}
    \frac{\mathcal{G}_f}{\ell} = \int_0^{\infty} \sigma(\varepsilon,k) \textrm{d} \varepsilon,
\end{equation}
wherein $\ell$ is the numerical length scale. The integral on the right-hand side must be broken into three parts, one for the elastic response, one for the plastic response, and one for the cracking response. During the elastic response we have
\begin{equation}
    \sigma = E\varepsilon^e,
\end{equation}
where $E$ is the elastic modulus of the material. As the elastic strain reaches the yield limit
\begin{equation}
    \varepsilon^e_y = \frac{\sigma_y}{E},
\end{equation}
the plastic strain $\varepsilon^p$ and damage index $d$ start growing. Hence, we have
\begin{equation} \label{eq:plasticResponse}
    \sigma = \exp(-\alpha k)\sigma_y.
\end{equation}
On the other hand, the expression in \Cref{eq:damageInternalPlastic} can be rewritten for this idealized one-dimensional case as follows
\begin{equation}
    \dot{k} = w(\tilde{\sigma}) \dot\varepsilon^p,
\end{equation}
wherein, according to \Cref{eq:weight}, we have 
\begin{equation}
    w(\tilde{\sigma}) = 1.
\end{equation}
As a result, the stress-strain response in \Cref{eq:plasticResponse} can be rewritten as
\begin{equation}
    \sigma = \exp(-\alpha \varepsilon^p)\sigma_y.
\end{equation}
In addition, the onset of cracking coincides with
\begin{equation}
     k_{c} = \varepsilon^p_{c},
\end{equation}
where $\varepsilon^p_{c}$ is the critical plastic strain. As a result, the damage internal variable after onset of cracking is defined for this idealized case as 
\begin{equation}
    k = \varepsilon^p_{c} + \varepsilon^d.
\end{equation}
Now, the cracking response is given by
\begin{equation}
    \sigma = \exp(-\alpha \varepsilon^p_{c} - \alpha \varepsilon^d) \sigma_y.
\end{equation}
Finally, the relation in \Cref{eq:fracture_energy} is rewritten as
\begin{align}
    \begin{split}
    \frac{\mathcal{G}_f}{\ell} &= \int_0^{\varepsilon_y} {E\varepsilon^e \textrm{d}\varepsilon^e} +
    \int_0^{\varepsilon^p_{c}} {\exp(-\alpha\varepsilon^p)\sigma_y \textrm{d}\varepsilon^p} \\
    &+ \int_0^{\infty} {\exp(-\alpha\varepsilon^p_{c} - \alpha\varepsilon^d)\sigma_y \textrm{d}\varepsilon^d}. \\
    \end{split}
\end{align}
Solving the definite integral gives
\begin{equation}
    \frac{\mathcal{G}_f}{\ell} = \frac{1}{2}\frac{\sigma _y^2}{E} + \frac{\sigma _y}{\alpha}.
\end{equation}
Finally, we arrive at
\begin{equation}
    \alpha = \frac{2E\ell\sigma_y}{2E\mathcal{G}_f-\ell\sigma_y^2}\label{eq:damageConstant}.
\end{equation}
It is worth mentioning that $\alpha$ must be positive, otherwise the damage index $d$ ranges over the interval $[0,-\infty]$, which is obviously incorrect. Since all the material parameters, including the fracture energy $\mathcal{G}_f$, the elastic modulus $E$, and the yield stress $\sigma_y$ are positive, the numerator is always larger than zero. Hence, the denominator must be positive to have a realistic damage growth. By setting
\begin{equation}
    2E\mathcal{G}_f-\ell\sigma_y^2 > 0,
\end{equation}
we arrive at
\begin{equation}
    \ell < \frac{2E\mathcal{G}_f}{\sigma_y^2},
\end{equation}
which defines the upper bound of the characteristic length. Hence, the domain of a problem must be discretized such that the numerical length scale does not exceed its upper bound.

\section{Implementation}\label{sec:implementation}
In order to make the discontinuous strain method as accessible as possible, we first present a one-dimensional implementation (\Cref{ssec:1D}) and subsequently elaborate on the differences to two and three dimensions (\Cref{ssec:2D+3D}). The implementation is presented independently of the underlying elasto-plastic-damage model. Thus any elasticity, plasticity, and damage model can be combined with the presented implementation.

\begin{algorithm*}[htb]
	\caption{Material routine with the main modifications to a conventional plastic-damage routine highlighted in red (\texttt{checkCrackOpening} from \Cref{alg:crackUpdate} and \texttt{checkCrackClosure} from \Cref{alg:crackCheck}). Illustrated for one dimension. Note, that the material properties are treated as global quantities within the presented algorithm to keep the function interfaces as simple as possible. A corresponding Python implementation is made available in~\cite{herrmann_discontinuous_2023}. 
	}\label{alg:MaterialRoutine}
	\begin{algorithmic}[1]
		\Require strain increment $\Delta \varepsilon$, history variables from last load increment $\tilde{\sigma}, \varepsilon^p, \varepsilon^d, k, d$, material properties $E, \nu, \sigma_y, \beta, d_c, \alpha$ (cf.~\Cref{eq:damageConstant}), $k_c$ (cf.~\Cref{eq:damageThreshold}), length scale $\ell$ 
		\If{{\color{red}$\varepsilon^d \neq 0$}} \Comment{simplified~\Cref{eq:crackOpeningStrain}}
		    \State $\Delta\varepsilon, \varepsilon^d, k\leftarrow${\color{red}\texttt{checkCrackClosure(}}$\Delta\varepsilon,\varepsilon^d,k${\color{red}\texttt{)}} \Comment{cf.~\Cref{alg:crackUpdate}}
		\EndIf
		\If{{\color{red}$\varepsilon^d = 0$}} \Comment{simplified~\Cref{eq:crackOpeningStrain}}
		    \State $\tilde{\sigma}\leftarrow$\texttt{stressIncrement(}$\tilde{\sigma},\Delta\varepsilon$\texttt{)}
		    \If{$f(\tilde{\sigma}) > 0$} \Comment{cf.~\Cref{eq:yieldSurface,eq:KKT1}}
		        \State $\tilde{\sigma}, \Delta\varepsilon^p\leftarrow$\texttt{returnMapping(}$\tilde{\sigma}$\texttt{)} \Comment{cf.~\Cref{eq:solutionPlasticStrainUpdate}}
		        \State $\varepsilon^p,\varepsilon^d,\tilde{\sigma},k\leftarrow${\color{red}\texttt{checkCrackOpening(}}$\Delta\varepsilon, \Delta\varepsilon^p, \varepsilon^p,\tilde{\sigma},k${\color{red}\texttt{)}} \Comment{cf.~\Cref{alg:crackCheck}}
		    \EndIf
	    \EndIf
	    \State $d\leftarrow$\texttt{updateDamage(}{\color{red}$k$}\texttt{)} \Comment{cf.~\Cref{eq:damageEvolution}}
	    \State \texttt{updateHistoryVariables(}$\tilde{\sigma}, \varepsilon^p, \varepsilon^d, k, d$\texttt{)}
        \State \Return total stress $\sigma$ \Comment{cf.~\Cref{eq:trueStress}} 
    \end{algorithmic}
\end{algorithm*}

\subsection{One Dimension}\label{ssec:1D}
Considering the method in one dimension\footnote{For the complete routine, the reader is referred to a basic one-dimensional Python implementation made available in~\cite{herrmann_discontinuous_2023}.} allows for several simplifications, enabled by working with scalar quantities instead of tensors for the stresses and strains. To incorporate the discontinuous strain method in a conventional elasto-plastic-damage routine, only two changes have to be undertaken, as highlighted in red in \Cref{alg:MaterialRoutine}:
\begin{itemize}
\item checking for crack closure,
\item checking for crack opening.
\end{itemize}

Before the stress state is altered through a stress increment it must be ensured, that the current discontinuity strain $\varepsilon^d$ is zero (line~4 of \Cref{alg:MaterialRoutine}). For a non-zero discontinuity strain, the strain increment $\Delta \varepsilon$ fully contributes to the discontinuity strain until the discontinuity strain becomes negative, hence indicating crack closure. This is therefore checked by \texttt{checkCrackClosure} (line~2 of \Cref{alg:MaterialRoutine}) presented in \Cref{alg:crackCheck}. If the discontinuity strain becomes negative, it is set to zero (line~4 of \Cref{alg:crackUpdate}), thus ensuring that the stress state is updated (lines~5-7 of \Cref{alg:MaterialRoutine}) --- when the main material routine checks for zero discontinuity strain (line~4 of \Cref{alg:MaterialRoutine}).

The stress is incremented with the strain increment $\Delta \varepsilon$ in line~5 of~\Cref{alg:MaterialRoutine}. In case the new stress state lies outside of the yield surface, i.e., $f(\tilde{\sigma}) > 0$, a new valid stress state is provided by the function \texttt{returnMapping} along with an increment of the plastic strain $\Delta\varepsilon^p$.


Subsequent to the return mapping, a potential crack opening, i.e., a contribution to the discontinuity strain $\varepsilon^d$ due to the excessive accumulation of plastic strain $\varepsilon^p$ and thereby damage $d$, is considered (line~8 of \Cref{alg:MaterialRoutine}) with \texttt{checkCrackOpening} from \Cref{alg:crackUpdate}. Exceeding the threshold of the damage internal variable $k_c$ defined in \Cref{eq:damageThreshold} leads to a crack opening and the total strain increment $\Delta \varepsilon$ is assigned to the (previously zero) discontinuity strain $\varepsilon^d$ (line~2 of \Cref{alg:crackUpdate}) instead of updating the plastic strain $\varepsilon^p$ (line~6 of \Cref{alg:crackUpdate}). In addition, during crack opening, the effective stress increment --- expressed in terms of total $\Delta \varepsilon$ and plastic strain increment $\Delta\varepsilon^p$ --- is removed from the altered effective stress $\tilde{\sigma}$ (line~3 of \Cref{alg:crackUpdate}) to restore the initial stress state prior to the applied material routine.

Note that the next time the material routine is called, the discontinuity strain $\varepsilon^d$ will be greater than zero, and \Cref{alg:crackCheck} will thereby be called in line~2 of \Cref{alg:MaterialRoutine}.

Common to both helper routines is that the damge internal variable $k$ (\Cref{eq:updateDamageInternalVariable}) is updated (line~6 of \Cref{alg:crackCheck} and lines~4 and~7 of \Cref{alg:crackUpdate}). 

Lastly, the discontinuity strain $\varepsilon^d$ is incorporated in the damage computation through the damage internal variable $k$ (line~11 of \Cref{alg:MaterialRoutine}) with \Cref{eq:damageEvolution} maintaining the damage evolution in the absence of plastic strain growth.


\begin{algorithm}[htb]
    \caption{\texttt{checkCrackClosure} for one dimension.}\label{alg:crackCheck}
    \begin{algorithmic}[1]
        \Require $\Delta \varepsilon, \varepsilon^d, k$
        \State $\varepsilon^d = \varepsilon^d + \Delta \varepsilon$
        \If{$\varepsilon^d < 0$} \Comment{check crack closure\\\hspace{4.9cm}cf.~\Cref{eq:crackOpeningStrain}}
            \State $\varepsilon^d = 0$ \Comment{deactivate $\varepsilon^d$ at crack closure}
        \Else
            \State $k = k + \langle \Delta \varepsilon \rangle$ \Comment{simplified~\Cref{eq:updateDamageInternalVariable}} 
        \EndIf
        \State \Return $\Delta\varepsilon, \varepsilon^d, k$
    \end{algorithmic}
\end{algorithm}

\begin{algorithm}[htb]
    \caption{\texttt{checkCrackOpening} for one dimension.}\label{alg:crackUpdate}
    \begin{algorithmic}[1]
        \Require $\Delta \varepsilon, \Delta \varepsilon^p, \varepsilon^p, \tilde{\sigma}, k$
        \If{$(k + \langle \Delta \varepsilon^p \rangle) > k_c$ and $\Delta \varepsilon > 0$}
            \State $\varepsilon^d = \Delta \varepsilon$
            \State $\tilde{\sigma} = \tilde{\sigma} - E(\Delta\varepsilon - \Delta\varepsilon^p)$ \Comment{reset stress}
            \State $k = k + \langle \Delta \varepsilon \rangle$ \Comment{simplified~\Cref{eq:damageInternalPlastic}}
        \Else
            \State $\varepsilon^p = \varepsilon^p + \Delta \varepsilon^p$
            \State $k = k + \langle \Delta \varepsilon^p \rangle$ \Comment{simplified~\Cref{eq:damageInternalPlastic}}
        \EndIf
        
        \State \Return $\varepsilon^p, \varepsilon^d, \tilde{\sigma}, k$
    \end{algorithmic}
\end{algorithm}

\subsection{Two \& Three Dimensions}\label{ssec:2D+3D}
The main differences to the one-dimensional implementations are, firstly, tensorial quantities instead of scalars for strains and stresses, and secondly, the determination of the crack surface orientation, quantified by its normal direction $\boldsymbol{n}$. The change, leads to minor modifications in two and three dimensions, highlighted in red in \Cref{alg:crackCheck3D} for the crack closure check and in \Cref{alg:crackUpdate3D} for the crack opening check. \Cref{alg:crackUpdate3D} now relies on the normal direction $\boldsymbol{n}$ to determine the crack opening strain using \Cref{eq:crackOpeningStrain} (line~1 of \Cref{alg:crackUpdate3D}). This is similarly employed for updating the damage internal variable $k$ when determining the crack closure in \Cref{alg:crackCheck3D} (lines~2 and 6). However, additionally the crack surface orientation must be determined using the maximum principal stress $\hat{\tilde{\sigma}}_\text{max}$ (line~7 in \Cref{alg:crackUpdate3D}). To ensure, that the normal direction $\boldsymbol{n}$ is only determined once, i.e., at the initial crack opening, a flag \texttt{crackInitiated} is employed (lines~5-6 in \Cref{alg:crackUpdate3D}).

The main material routine from \Cref{alg:MaterialRoutine} remains largely unchanged apart from the dimensionality of the stresses and strains and the crack opening strain checks in lines~1 and~4, which now rely on the crack normal $\boldsymbol{n}$ using \Cref{eq:crackOpeningStrain}. However, the two-dimensional plane stress version requires further algorithmic manipulations to ensure that the return mapping fulfills a zero out-of-plane stress state. The procedure is summarized in \Cref{sec:planeStressStrain}.

\begin{algorithm}[htb]
    \caption{\texttt{checkCrackClosure} for two \& three dimensions. Changes to one-dimensional implementation from \Cref{alg:crackCheck} marked in red.}\label{alg:crackCheck3D}
    \begin{algorithmic}[1]
        \Require $\Delta \boldsymbol{\varepsilon}, \boldsymbol{\varepsilon}^d, k$
        \State $\boldsymbol{\varepsilon}^d = \boldsymbol{\varepsilon}^d + \Delta \boldsymbol{\varepsilon}$    
        \If{{\color{red}$\boldsymbol{n}^\intercal\boldsymbol{\varepsilon}^d\boldsymbol{n} < 0$}}  \Comment{check crack closure\\\hspace{4.9cm}cf.~\Cref{eq:crackOpeningStrain}}
            \State $\boldsymbol{\varepsilon}^d = \boldsymbol{0}$ \Comment{deactivate $\boldsymbol{\varepsilon}^d$ at crack closure}
        \Else
            \State $k = k$ $+$ {\color{red} $\langle\boldsymbol{n}^\intercal\Delta \boldsymbol{\varepsilon}\boldsymbol{n}\rangle$}  \Comment{cf.~\Cref{eq:updateDamageInternalVariable}}
        \EndIf
        \State \Return $\Delta\boldsymbol{\varepsilon}, \boldsymbol{\varepsilon}^d, k$
    \end{algorithmic}
\end{algorithm}

\begin{algorithm}[htb]
    \caption{\texttt{checkCrackOpening} for two \& three dimensions. Changes to one-dimensional implementation from~\Cref{alg:crackUpdate} marked in red.}\label{alg:crackUpdate3D}
    \begin{algorithmic}[1]
        \Require $\Delta \boldsymbol{\varepsilon}, \Delta \boldsymbol{\varepsilon}^p, \boldsymbol{\varepsilon}^p, \tilde{\boldsymbol{\sigma}}, k$
        
        \If{$(k$ $+$ {\color{red}$w(\tilde{\boldsymbol{\sigma}})\Delta\varepsilon^p_\text{max}$}$)>k_c$ and {\color{red}$\boldsymbol{n}^\intercal\Delta\boldsymbol{\varepsilon}\boldsymbol{n}$} $> 0$}
            \State $\boldsymbol{\varepsilon}^d=\Delta\boldsymbol{\varepsilon}$
            \State $\tilde{\boldsymbol{\sigma}} = \tilde{\boldsymbol{\sigma}} - \tilde{\boldsymbol{D}}(\Delta\boldsymbol{\varepsilon}-\Delta\boldsymbol{\varepsilon}^p)$ \Comment{reset stress}
            \State $k = k + $ {\color{red}$w(\tilde{\boldsymbol{\sigma}})\Delta\varepsilon_\text{max}$} \Comment{cf.~\Cref{eq:damageInternalPlastic}}
            \If{{\color{red}\texttt{crackInitiated}$=$\texttt{False}}}
                \State {\color{red}\texttt{crackInitiated}$=$\texttt{True}}
                \State {\color{red}$\boldsymbol{n}\leftarrow$\texttt{getCrackNormal(}$\tilde{\boldsymbol{\sigma}}$\texttt{)}} \Comment{using $\hat{\tilde{\sigma}}_\text{max}$}
            \EndIf
        \Else 
            \State $\boldsymbol{\varepsilon}^p=\boldsymbol{\varepsilon}^p + \Delta\boldsymbol{\varepsilon}^p$ 
            \State $k = k + $ {\color{red}$w(\tilde{\boldsymbol{\sigma}})\Delta\varepsilon^p_\text{max}$} \Comment{cf.~\Cref{eq:damageInternalPlastic}}
        \EndIf
        \State \Return $\boldsymbol{\varepsilon}^p, \boldsymbol{\varepsilon}^d, \tilde{\boldsymbol{\sigma}}, k$, {\color{red}$\boldsymbol{n}$, \texttt{crackInitiated}}
    \end{algorithmic}
\end{algorithm}

\section{Numerical Results}\label{sec:results}
To emphasize the purpose of the proposed method, a one-dimensional example (\Cref{ssec:Results1D}) is considered showcasing the error incurred during reloading when neglecting the crack closure. Furthermore, the method is validated against established benchmarks from the literature (\Cref{ssec:Results2D}). A center-notched beam (\Cref{sssec:Results2Da}), an off center-notched beam (\Cref{sssec:Results2Db}), and a L-shaped panel (\Cref{sssec:Results2Dc}) are considered in two dimensions. The corresponding material properties are summarized in \Cref{tab:material}. The implementations are provided in~\cite{herrmann_discontinuous_2023} in form of a Python code for the one-dimensional case, and in terms of an Abaqus C++ user material subroutine (UMAT) and driver files for the two-dimensional cases.

\begin{table}[htb]
    \centering
    \caption{Material parameters.}\label{tab:material}
    \setlength{\tabcolsep}{3pt}
    \begin{tabular}{@{}cccc@{}}
    \hline
           & \begin{tabular}{c}Center-\\notched\\beam~\cite{schroder_phase-field_2022}\end{tabular} 
           & \begin{tabular}{c}Off center-\\notched\\beam~\cite{jenq_mixed-mode_1988}\end{tabular}   
           & \begin{tabular}{c}L-\\shaped\\panel~\cite{winkler_traglastuntersuchungen_2001}\end{tabular}  \\
    \hline
    $E \textrm{ [GPa]}$             & 54    & 34   & 22.5  \\
    $\nu \textrm{ [--]}$           & 0.2   & 0.2  & 0.2   \\
    $\sigma_y \textrm{ [MPa]}$      & 7.2   & 4    & 2.3   \\
    $\mathcal{G}_f \textrm{ [MPa$\cdot$mm]}$ & 0.075 & 0.09 & 0.09 \\
    $\beta \textrm{ [--]}$         & 0.2   & 0.2  & 0.2   \\
    $d_c \textrm{ [--]}$           & 0.35  & 0.4  & 0.4   \\
    \hline
    \end{tabular}
\end{table}

\subsection{One Dimension}\label{ssec:Results1D}
\begin{figure}[htb]
	\centering
	\begin{subfigure}[b]{0.49\textwidth}
		\includegraphics[width=\textwidth]{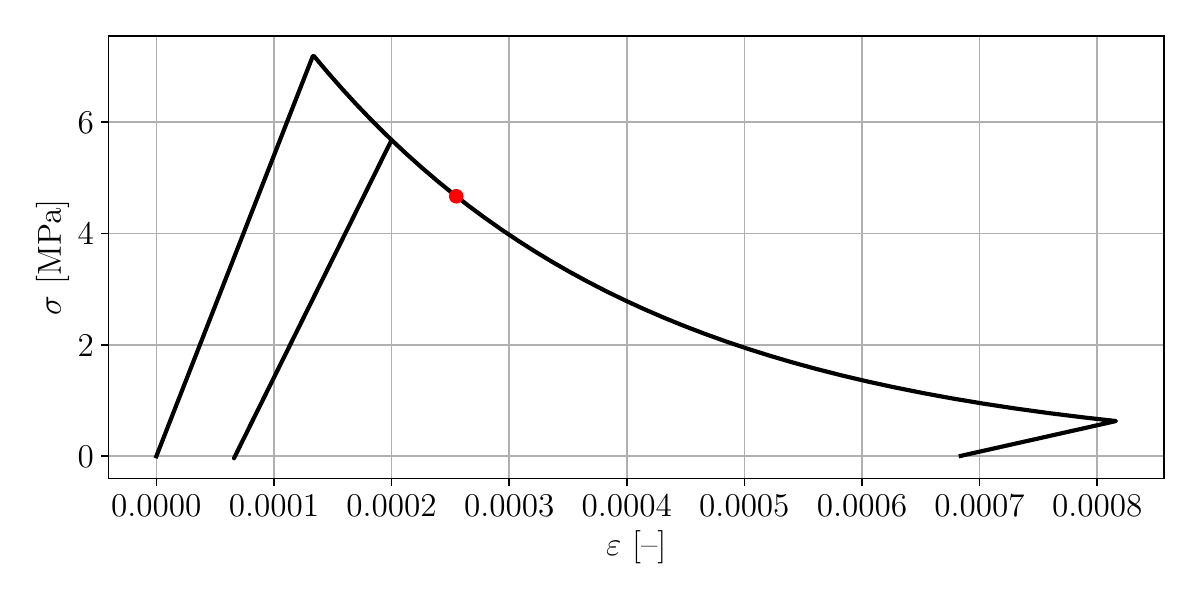}
		\caption{without discontinuity strain.}\label{fig:1DWithout}
	\end{subfigure}
	\hfill
	\begin{subfigure}[b]{0.49\textwidth}
	    \begin{tikzpicture}
		\node(picA){\includegraphics[width=\textwidth]{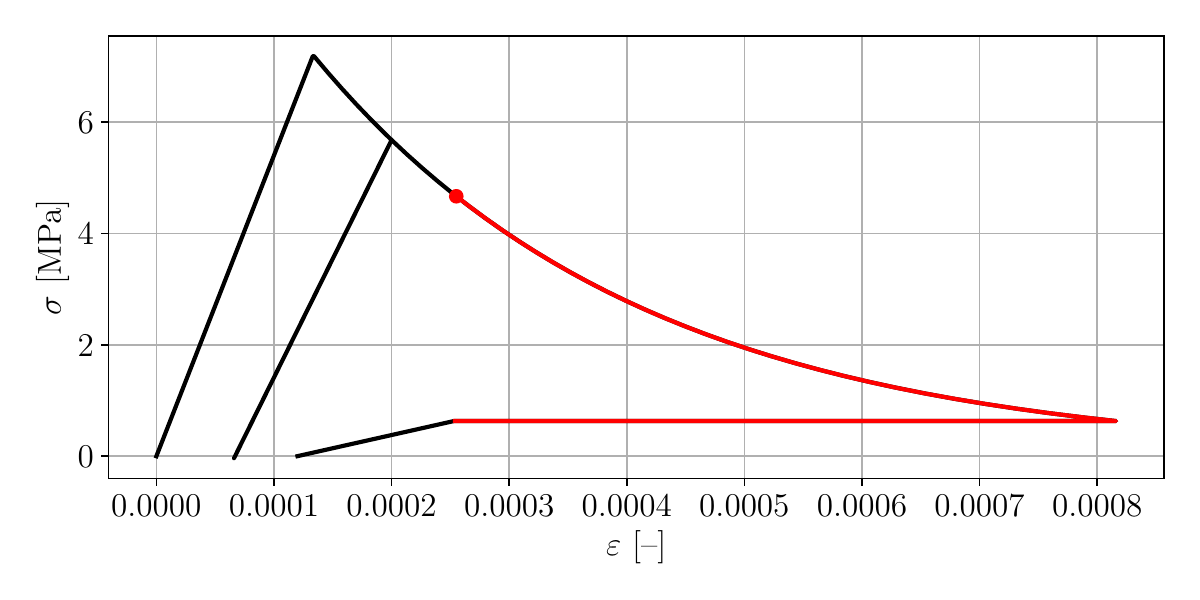}};
		\draw [decorate,decoration = {brace},thick, red] (-1.07,-0.7) -- (3.8,-0.7);
		\node [red] at (1.45,-0.12) {$\varepsilon^d$};
		\end{tikzpicture}
		\caption{with discontinuity strain.}\label{fig:1DWith}
	\end{subfigure}
	\caption{Stress-strain curve without and with discontinuity strain $\varepsilon^d$. The red circle indicates that critical damage threshold has been breached. The material parameters employed to obtain these results are from the center-notched beam (\Cref{tab:material}) with a length scale of $\ell=30$ mm.}\label{fig:1Da}
\end{figure}

\begin{figure}[htb]
\centering
	\begin{subfigure}[b]{0.49\textwidth}
		\includegraphics[width=\textwidth]{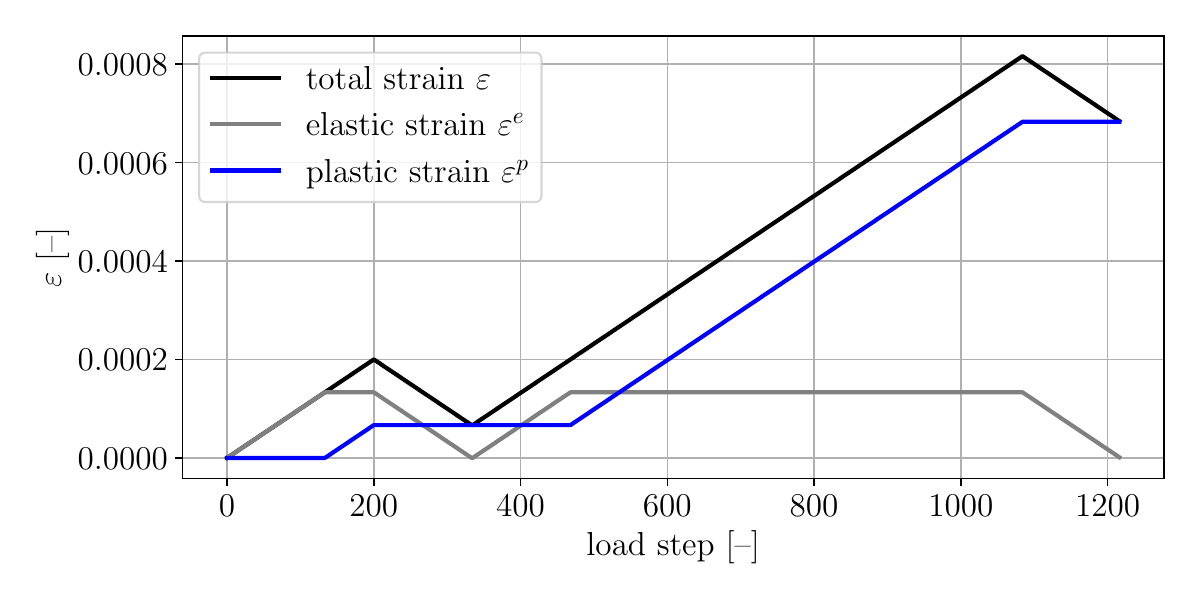}
		\caption{without discontinuity strain.}\label{fig:1DStrainWithout}
	\end{subfigure}
	\hfill
	\begin{subfigure}[b]{0.49\textwidth}
		\includegraphics[width=\textwidth]{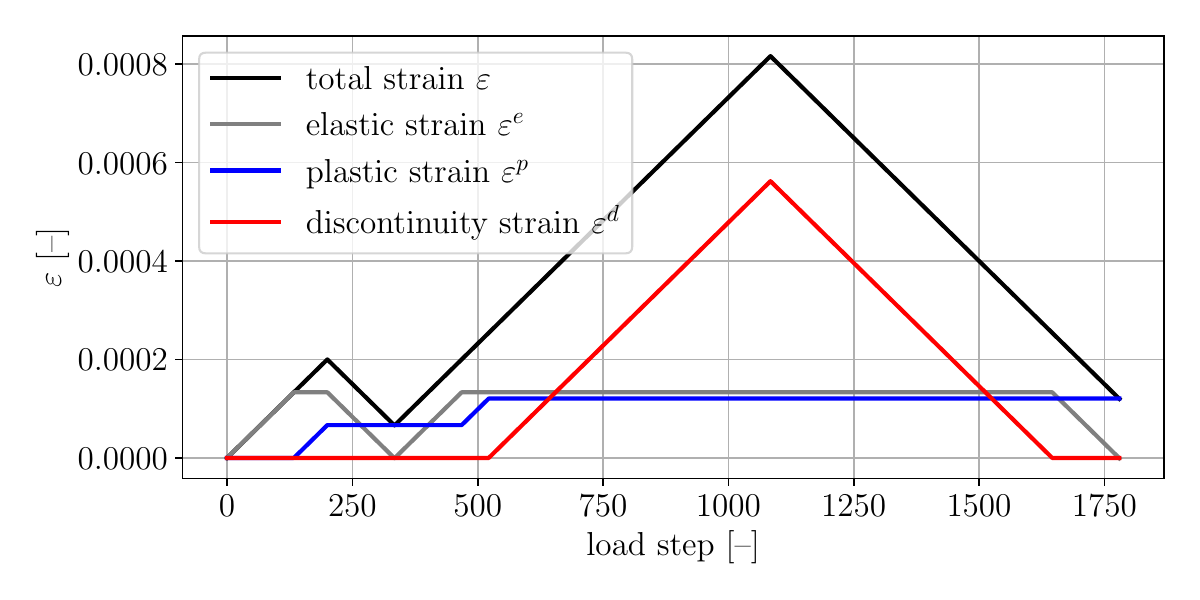}
		\caption{with discontinuity strain.}\label{fig:1DStrainWith}
	\end{subfigure}
	\caption{Strain evolution without and with discontinuity strain $\varepsilon^d$ corresponding to the stress-strain curves from \Cref{fig:1Da}.}\label{fig:1Db}
\end{figure}

A two-cycle displacement-controlled loading is imposed and illustrated in \Cref{fig:1Da,fig:1Db}. The first loading-unloading cycle is below the critical damage threshold $d_{c}$, while the second cycle is above. This yields significant differences in the stress-strain curves without (\Cref{fig:1DWithout}) and with (\Cref{fig:1DWith}) the discontinuity strain $\varepsilon^d$. During the first loading and unloading, the behavior is identical. However, the responses diverge as soon as the critical damage threshold $d_c$ is reached (indicated by the red circle in \Cref{fig:1DWithout,fig:1DWith}). Without the discontinuity strain, the irreversible plastic strain $\varepsilon^p$ continues accumulating (see \Cref{fig:1DStrainWithout}) despite the complete failure of the material point. With the discontinuity strain, the excess strain is absorbed by the reversible discontinuity strain $\varepsilon^d$ (indicated in red in \Cref{fig:1DWith} and visible only in \Cref{fig:1DStrainWith}). The stress-strain curves are indistinguishable until unloading after failure, i.e., until the crack closure arises. It only becomes apparent during unloading, where the excess irreversible plastic strain leads to an unphysical unloading path (\Cref{fig:1DWithout}). The correct path with no change in stresses until complete crack closure is depicted in \Cref{fig:1DWith} --- achieved through the discontinuity strain $\varepsilon^d$.

\subsection{Two Dimensions}\label{ssec:Results2D}
Now that the motivation of the proposed method is clarified with a basic one-dimensional example, the validity of the discontinuous strain method is assessed with three two-dimensional displacement-controlled cases relying on experimental results from the literature. Each of these cases serves a different purpose. The first, the center-notched beam (\Cref{sssec:Results2Da}) experiences pure mode I fracture, while the second, the off center-notched beam (\Cref{sssec:Results2Db}) exhibits a mixed-mode fracture consisting of mode I and mode II. Lastly, the L-shaped panel (\Cref{sssec:Results2Dc}) allows a loading under both tension and compression --- unlike the center- and off center-notched beams, where a tensile force $F$ would lead to failure in the vicinity of the applied load $F$. All of the considered cases are modeled as plane stress.

Both the effect of the discontinuity strain $\boldsymbol{\varepsilon}^d$ and the mesh independence are illustrated in the following (\Cref{sssec:Results2Da,sssec:Results2Db,sssec:Results2Dc}). Lastly, the impact of the introduced field $\boldsymbol{\varepsilon}^d$ on the convergence of the global Newton--Raphson scheme within the nonlinear finite element framework is discussed in the context of the computational effort (\Cref{ssec:compeffort}).

\subsubsection{Center-Notched Beam}\label{sssec:Results2Da}
\begin{figure}[htb]
    \centering
    \includegraphics[scale=1.0]{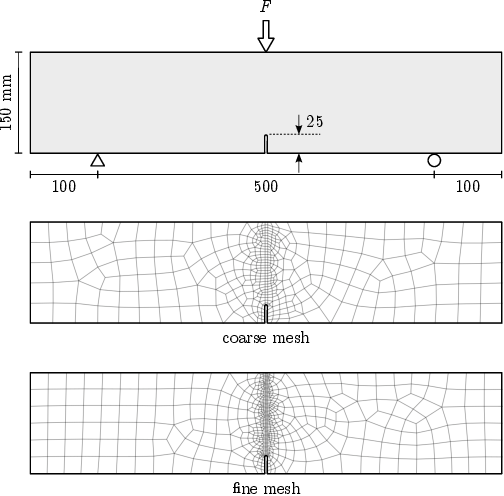}
    \caption{Geometry, boundary conditions, and finite element meshes of the center-notched beam.}
    \label{fig:center_geometry}
\end{figure}

The center-notched beam under three-point bending from~\cite{schroder_phase-field_2022} is considered first and depicted in \Cref{fig:center_geometry} showing the dimensions and boundary conditions. Under compression, a pure mode I crack initiating at the centered notch and propagating straight through the height of the beam is expected. The material properties presented in \Cref{tab:material} have been calibrated to the experimental data from~\cite{schroder_phase-field_2022}. The two finite element meshes employed for the simulation are also depicted in \Cref{fig:center_geometry}, which are refined towards a vertical line originating at the notch tip in anticipation of the crack evolution. A coarse and a fine mesh are considered in order to show the result's insensitivity to the discretization. 

\begin{figure}[htb]
	\centering
	\begin{subfigure}[b]{0.49\textwidth}
		\includegraphics[width=\textwidth]{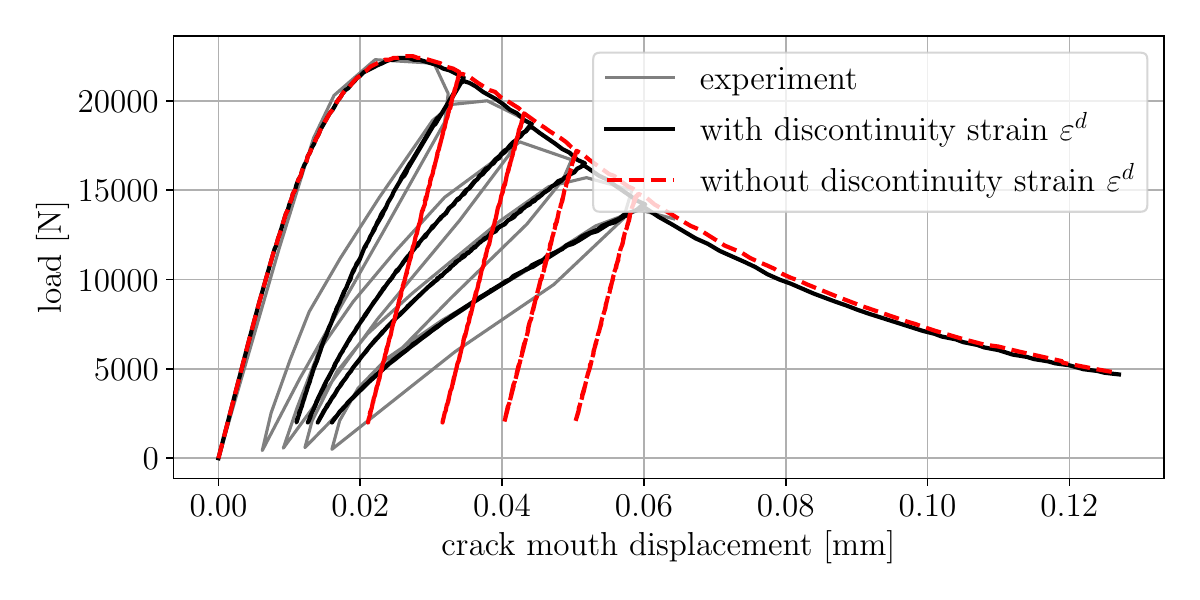}
		\caption{Influence of strain discontinuity $\varepsilon^d$.}\label{fig:case1Result1}
	\end{subfigure}
	\hfill
		\begin{subfigure}[b]{0.49\textwidth}
		\includegraphics[width=\textwidth]{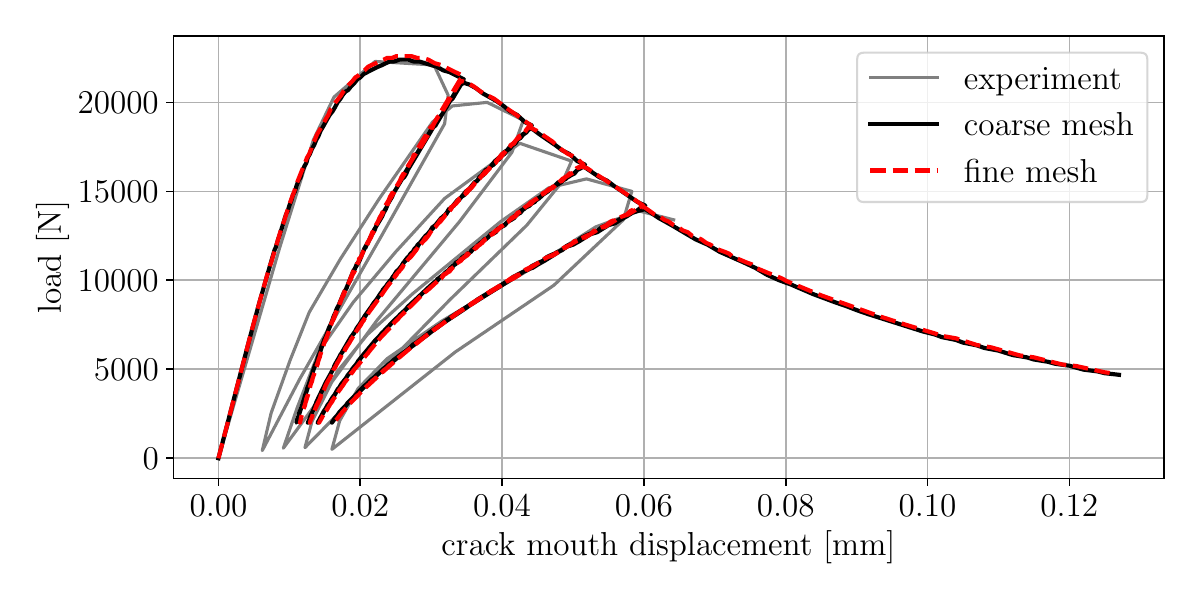}
		\caption{Mesh Sensitivity.}\label{fig:case1Result2}
	\end{subfigure}
	\caption{Stress strain curve of center-notched beam.}\label{fig:case1Result}
\end{figure}

The beam is subjected to cyclic loading in order to showcase the crack closure effect neglected by conventional ductile models. The applied load $F$ versus crack mouth displacement, i.e., opening at the bottom of the notch (see \Cref{fig:center_geometry}) is depicted in \Cref{fig:case1Result} highlighting the cyclic loading. \Cref{fig:case1Result1} compares the model with and without discontinuity strain $\varepsilon^d$, where it is visible that the model without discontinuity strain, i.e., the conventional plastic-damage model captures the unloading and reloading incorrectly --- seen when compared to the experimental curves in gray. The slopes remain almost as steep as during the initial loading despite a crack having formed, which should reduce the stiffness during unloading and reloading. This misrepresentation is caused by the continued accumulation of plastic strain within the crack. By contrast, the discontinuous strain method captures this effect correctly, as seen by the change in the slopes and the experimental agreement indicated in gray. Notice again that the difference only becomes apparent after unloading, as in the one-dimensional example from \Cref{ssec:Results1D}.

\begin{figure}[htb]
    \centering
    \includegraphics[scale=1.0]{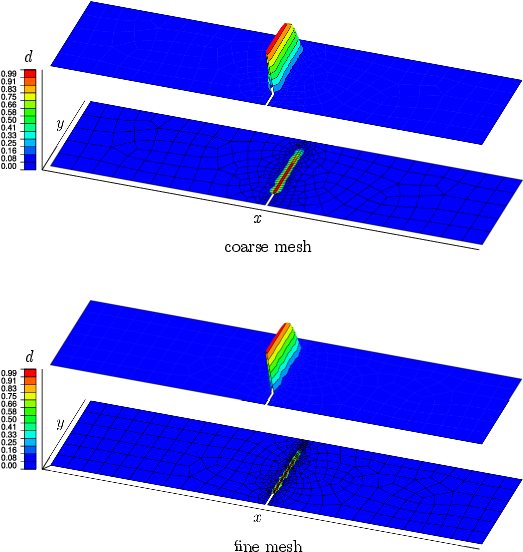}
    \caption{Damage contours of the center-notched beam.}
    \label{fig:center_contour}
\end{figure}

Additionally, the effect of the introduced characteristic length scale parameter $\ell$ from \Cref{ssec:energyEquivalence} is shown in \Cref{fig:case1Result2}, where no difference between the coarse and fine mesh can be noticed. This is similarly seen in the resulting cracks quantified by the damage variable $d$, see \Cref{fig:center_contour}. Apart from the width of the fracture process zone (corresponding to one element), the crack has evolved in the same manner. 

\begin{figure}[htb]
	\centering
	\begin{subfigure}[b]{0.49\textwidth}
		\includegraphics[width=\textwidth]{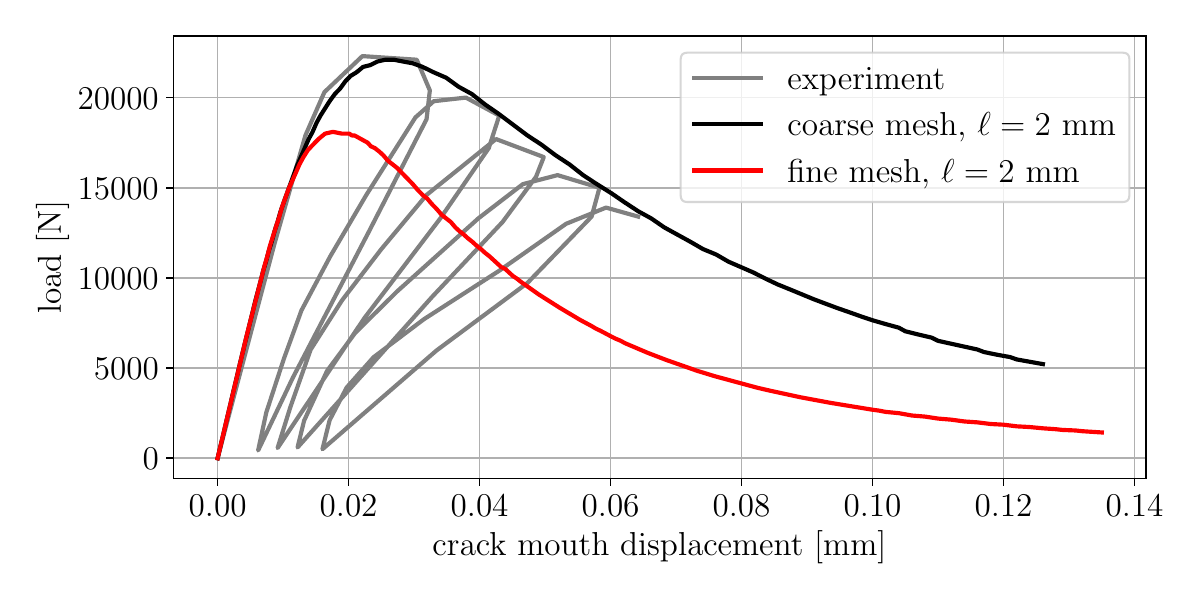}
	\end{subfigure}
	\caption{Demonstration of the influence of the discretization's length scale $\ell$ without the energy equivalence derived in \Cref{ssec:energyEquivalence}. The length scale $\ell=2$ mm is applied to all elements within the coarse and fine mesh, resulting in major differences in the global response between the two.}\label{fig:case1ellDependency}
\end{figure}

By contrast, \Cref{fig:case1ellDependency} highlights the severe differences in the global response if the mesh dependent effects are neglected by not introducing a length scale $\ell$. To showcase the problem, the same length scale is applied across all finite elements, independent of their true sizes. This global length scale is calibrated to $\ell=2$ mm, such that the coarse mesh result matches the experimental curves in \Cref{fig:case1ellDependency}. Utilizing the same length scale for the fine mesh leads to a significantly different response. This is because less energy is dissipated when utilizing the finer mesh with the length scale $\ell=2$ mm, seen in terms of an earlier crack formation. Elements with a too big length scale $\ell$ have a larger damage growth constant $\alpha$ (\Cref{eq:damageConstant}), leading to faster damage accumulation (\Cref{eq:plasticResponse}). Thus, the method becomes mesh dependent if the length scale explained in \Cref{ssec:energyEquivalence} is not employed correctly.

\subsubsection{Off Center-Notched Beam}\label{sssec:Results2Db}
\begin{figure}[htb]
    \centering
    \includegraphics[scale=1.0]{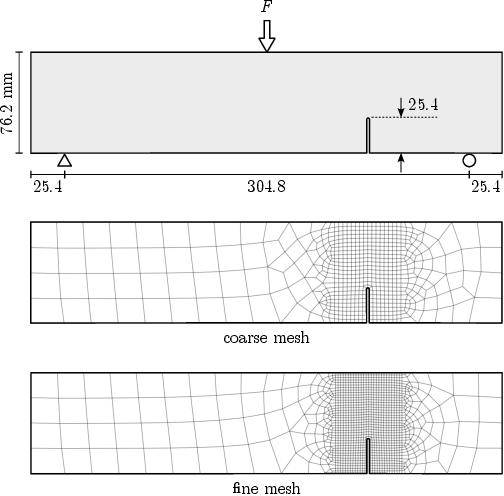}
    \caption{Geometry, boundary conditions, and finite element meshes of the off center-notched beam.}
    \label{fig:off_geo}
\end{figure}

Offsetting the notch of the beam results in a mixed-mode condition composed of mode I and mode II, due to the shear loads at the notch tip. Specifically, the experiment by~\cite{jenq_mixed-mode_1988} is considered and illustrated with dimensions and boundary conditions in \Cref{fig:off_geo}. For the calibrated material parameters, see \Cref{tab:material}. A mode II crack is expected to originate at the notch tip until a mode I crack takes over, propagating vertically through the height of the beam. Anticipating this fracture pattern, the mesh is refined within the domain of the expected crack in a coarse mesh and a fine mesh, also depicted in \Cref{fig:off_geo}.

\begin{figure}[htb]
	\centering
	\begin{subfigure}[b]{0.49\textwidth}
		\includegraphics[width=\textwidth]{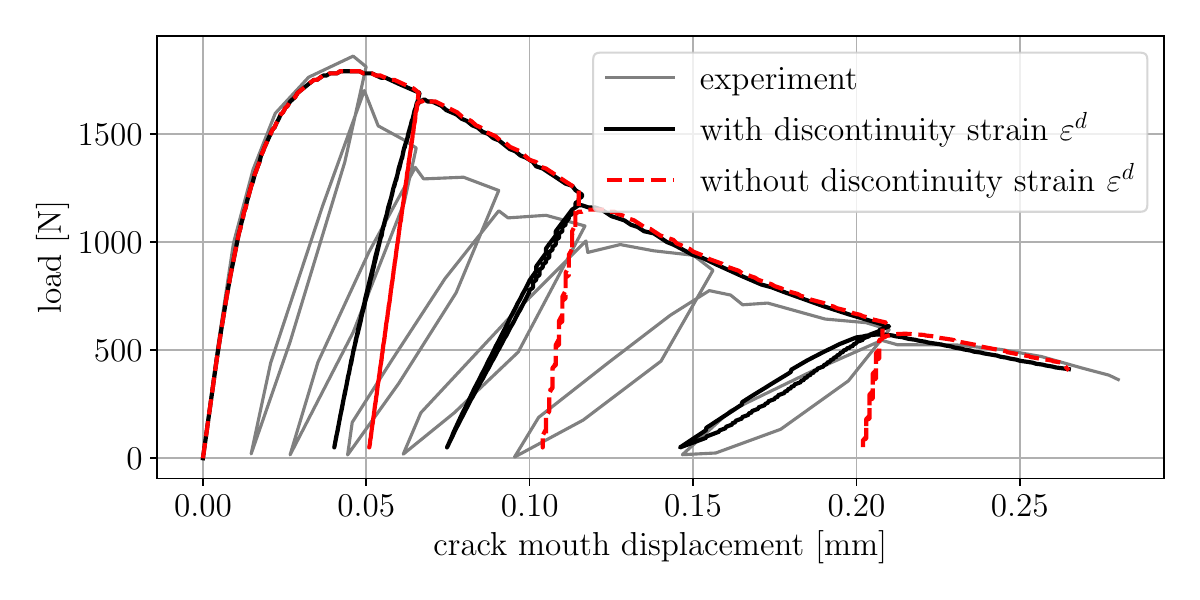}
		\caption{Influence of strain discontinuity $\varepsilon^d$.}\label{fig:case2Result1}
	\end{subfigure}
	\hfill
		\begin{subfigure}[b]{0.49\textwidth}
		\includegraphics[width=\textwidth]{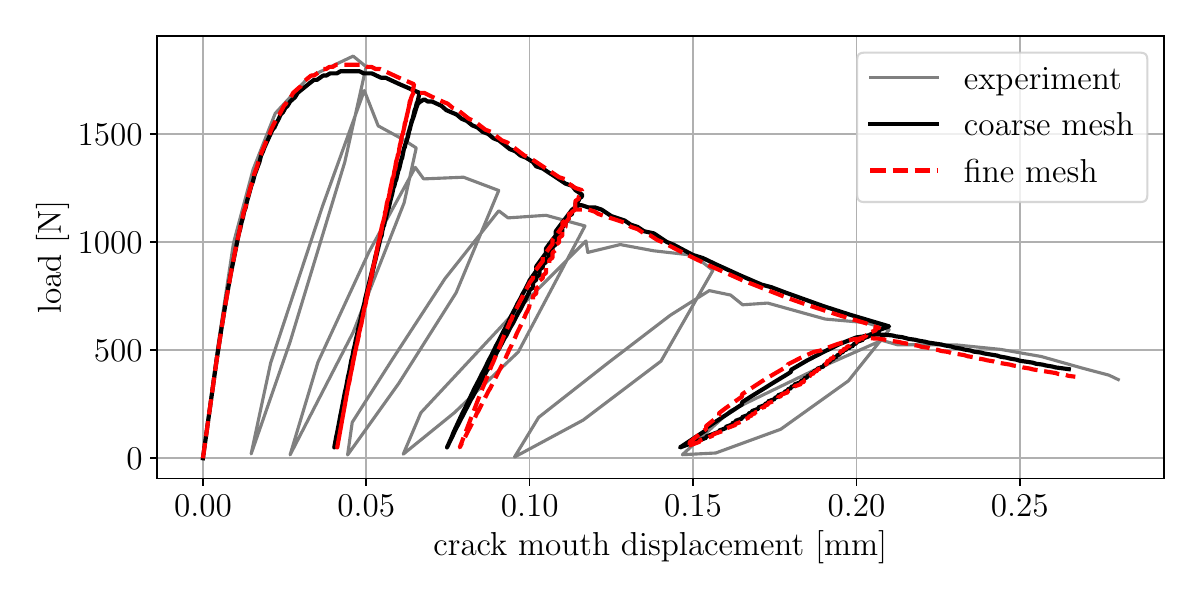}
		\caption{Mesh Sensitivity.}\label{fig:case2Result2}
	\end{subfigure}
	\caption{Stress strain curve of off center-notched beam.}\label{fig:case2Result}
\end{figure}

\begin{figure}[htb]
    \centering
    \includegraphics[scale=1.0]{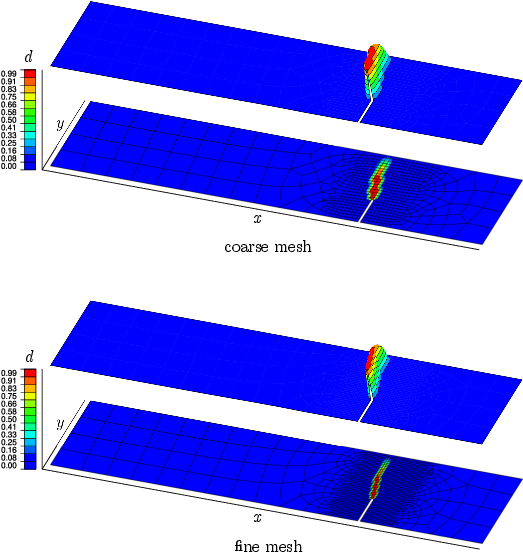}
    \caption{Damage contours of the off center-notched beam.}
    \label{fig:off_contour}
\end{figure}

As similarly shown for the center--notched beam, the loading curves in \Cref{fig:case2Result1} begin to deviate during unloading within a cyclic loading. Again, the discontinuous strain method accurately represents the experimental results from~\cite{jenq_mixed-mode_1988}, whereas not incorporating the discontinuity strain yields artificially stiffened unloading/reloading paths. \Cref{fig:case2Result2} compares the coarse and fine mesh results with only minor differences. By contrast to the centered-notched beam mesh sensitivity study from \Cref{fig:case1Result2}, minor deviations were observed --- most likely caused by the mode II fracture. The overall curves, however, still show the same tendencies, thus illustrating the mesh independence warranted by the length scale parameter $\ell$. This is similarly reflected by the contour plots of the damage variable in \Cref{fig:off_contour}, where the initial skewed mode II crack can be distinguished from the vertically growing mode I crack. Again, no difference in pattern can be identified except for the fracture process zone width --- composed of one element.

\subsubsection{L-Shaped Panel}\label{sssec:Results2Dc}

\begin{figure*}[htb]
    \centering
    \includegraphics[scale=1.0]{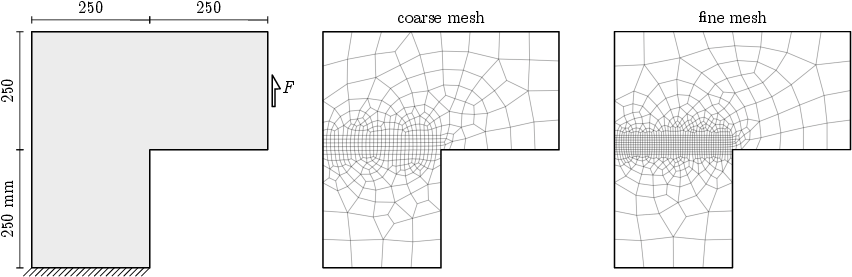}
    \caption{Geometry, boundary conditions, and finite element meshes of the L-shaped panel.}
    \label{fig:ell_geo}
\end{figure*}

Lastly, the L-shaped panel from~\cite{winkler_traglastuntersuchungen_2001} is considered. The geometry and boundary conditions are provided in \Cref{fig:ell_geo}, while the calibrated material properties are given by \Cref{tab:material}. The crack nucleation is anticipated at the singularity caused by the re-entrant corner and propagated horizontally to the left edge. The applied shear load $F$ allows for both positive and negative loads, i.e., both tensile and compressive forces on the crack interface. Again, a coarse and fine mesh are employed in order to show the mesh insensitivity. These are depicted in \Cref{fig:ell_geo}. 

\begin{figure}[htb]
	\centering
	\begin{subfigure}[b]{0.49\textwidth}
		\includegraphics[width=\textwidth]{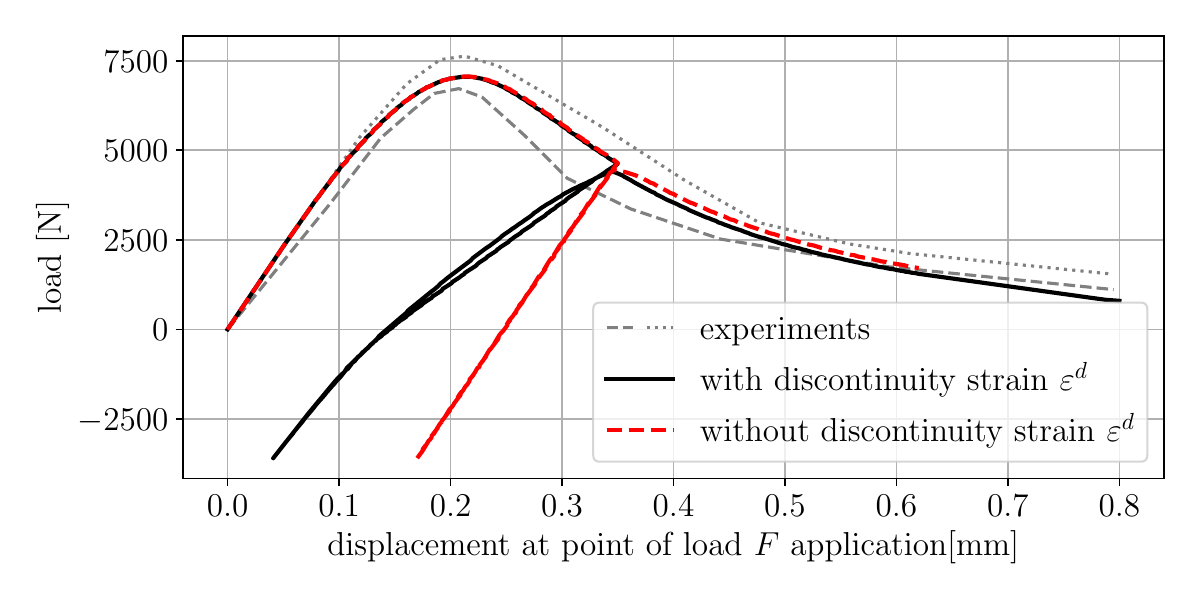}
		\caption{Influence of strain discontinuity $\varepsilon^d$.}\label{fig:case3Result1}
	\end{subfigure}
	\hfill
		\begin{subfigure}[b]{0.49\textwidth}
		\includegraphics[width=\textwidth]{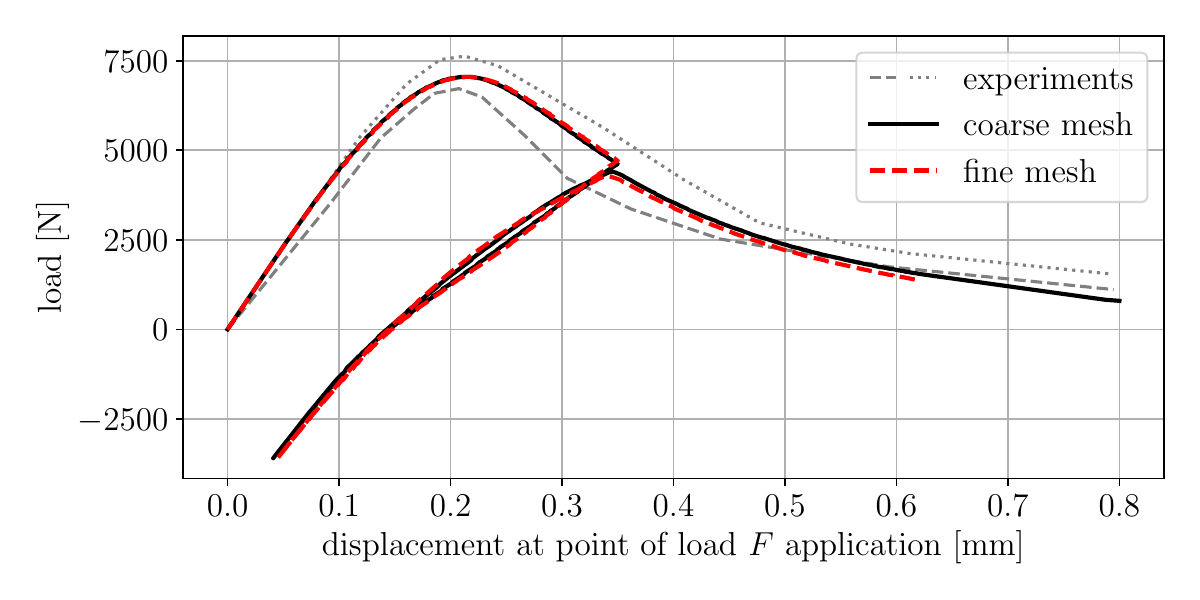}
		\caption{Mesh Sensitivity.}\label{fig:case3Result2}
	\end{subfigure}
	\caption{Stress strain curve of L-shaped panel.}\label{fig:case3Result}
\end{figure}

As in the previous cases, the structure is subjected to cyclic loading. The corresponding response is depicted in \Cref{fig:case3Result1}, where the main difference from the earlier cases is the compressive load acting during the unloading. This emphasizes the difference between the slopes of the model without and with discontinuity strain $\varepsilon^d$. Also, as in the first two examples, the mesh independency of the results enforced by the length scale $\ell$ is successful, as seen in \Cref{fig:case3Result}. The corresponding damage contour plots are provided in \Cref{fig:off_contour}, showing the horizontal fracture with a minor initial mode II crack.

\begin{figure*}[htb]
    \centering
    \includegraphics[scale=1.0]{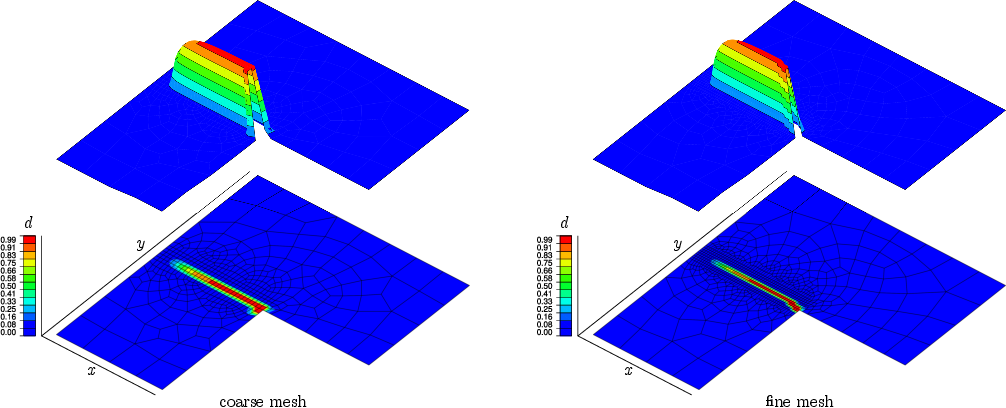}
    \caption{Damage contours of the L-shaped panel.}
    \label{fig:ell_contour}
\end{figure*}


\subsubsection{Computational Effort}\label{ssec:compeffort}
Finally, the question of additional computational effort due to the introduced discontinuity strain $\varepsilon^d$ arises. Most important is the impact on the convergence of the global Newton--Raphson solver. This is determined by the total number of Newton--Raphson iterations --- including iterations contributing to a load step with failed convergence leading to another load step with a reduced load increment. If one considers the three cases with and without discontinuity strain $\varepsilon^d$, as previously shown in \Cref{fig:case1Result1,fig:case2Result1,fig:case3Result1}, the relative increase in total Newton--Raphson iterations is 
\begin{itemize}
\begin{samepage}
    \item $6.45\%$ for the center-notched beam,
    \item $0.54\%$ for the off center-notched beam,
    \item $4.70\%$ for the L-shaped panel.
\end{samepage}
\end{itemize}
Thus, the effect on the global convergence is negligible. What is, however, not entirely negligible is the memory footprint caused by the additional history variables per integration point: $\boldsymbol{\varepsilon}^d$ (\texttt{double}$\times N(N+1)/2$), $\boldsymbol{n}$ (\texttt{double}$\times N$), \texttt{crackInitiated} (\texttt{bool}). The memory requirement of the history variables thereby increases by a factor of $\sim 1.64$ in two dimensions ($N=2$) and a factor of $\sim 1.65$ in three dimensions $(N=3)$.



\section{Conclusion}\label{sec:conclusion}
Currently, most ductile models do not correctly consider the deformations after crack formation, i.e., they contribute to the plastic strain and are thus irrecoverable. This becomes apparent when the crack closes and reopens, e.g., in cyclic loading for low-cycle fatigue simulations. To this end, we propose the discontinuous strain method in which we employ an additional term, the discontinuity strain, as an extension of the additive strain decomposition. The discontinuity strain's purpose is to absorb the excess strain after failure and release it upon unloading. As the crack closure can be identified at the point where the discontinuity strain becomes zero, this extension allows for an accurate representation of the crack closure and reopening, i.e., during unloading and reloading after failure. The additional strain field only comes at a minor increase in computational effort, without drastically increasing the number of Newton--Raphson iterations, yet an increase in memory footprint by a factor of approximately $1.6$. Mesh independence is incorporated through a characteristic length scale connecting the discretization resolution to the crack band width through an energy equivalence.

We have demonstrated the method on a simple plasticity model for illustration purposes. However, the discontinuous strain method can readily be adapted to any other elasto-plastic-damage model.
\bigskip

\section*{Data Availability}
We provide implementations for the one- and two/three-dimensional cases in~\cite{herrmann_discontinuous_2023}. The one-dimensional version is made available as Python code, while the two/three-dimensional implementation consists of a C++ user material subroutine (UMAT) for Abaqus with corresponding driver files for the presented examples.

\section*{Acknowledgment}
The work was supported by the Alexander von Humboldt Foundation, the Geothermal-Alliance Bavaria (GAB) by the Bavarian State Ministry of Science and the Arts (StMWK) and Deutsche Forschungsgemeinschaft (DFG, Germany) through the project 414265976 TRR 277 C-01.

\appendix
\renewcommand\thesection{\Alph{section}}

\section{Special Treatment for Plane Stress}\label{sec:planeStressStrain}
Plane strain and axisymmetric cases are special forms of the general three-dimensional formulation in which some components of the total strain tensor vanish. Hence, they are simply attainable in a displacement-based formulation. On the other hand, the plane stress case relies on the fact that the out-of-plane components of the stress tensor are zero. Although this constraint is easily enforced in linear elasticity by redefining the constitutive matrix, plastic straining violates it since the normal to the flow potential is a vector with arbitrary direction in the Haigh--Westergaard stress space. Thus, the return-mapping procedure must be modified such that the updated stress tensor satisfies the plane stress conditions. For this purpose, we use a nested iteration procedure in which, first, the general three-dimensional return-mapping procedure is performed. Subsequently, the out-of-plane component of the strain increment is corrected within a modified Newton strategy such that its corresponding stress component becomes zero. To this end, we perform the following update
\begin{equation}
   \Delta\varepsilon_{33} := \Delta\varepsilon_{33} - \frac{\tilde\sigma_{33}}{\tilde{D}_{33}}
\end{equation}
and reperform the return-mapping with the new strain increment tensor. This process continues until
\begin{equation}
   |\tilde\sigma_{33}| \leq \epsilon_{tol},
\end{equation}
meaning that the magnitude of the out-of-plane component of the effective stress tensor falls below an allowable tolerance~\cite{de_souza_neto_computational_2011}.

\bibliographystyle{ieeetr}

\setlength{\bibsep}{3pt}
\setlength{\bibhang}{0.75cm}{\fontsize{9}{9}\selectfont\bibliography{library}}

\end{document}

%% file: configuration.tex
\usepackage{xcolor} 
\definecolor{lightblue}{rgb}{0,0.5,1.0}
\definecolor{linkblue}{rgb}{0,0.1,0.6}
\definecolor{citegreen}{rgb}{0,0.4,0.0}
\definecolor{linkred}{rgb}{0.8,0,0.005}
\definecolor{mailviolet}{rgb}{0.3,0,0.35}
\definecolor{tumblue}{rgb}{0,0.396,0.741}
\definecolor{darkgreen}{rgb}{0,0.4,0} 
\definecolor{darkbrown}{rgb}{0.5, 0.396, 0.09}

\usepackage[hyperindex,colorlinks=true,linkcolor=linkblue,citecolor=citegreen,urlcolor=mailviolet,filecolor=linkred]{hyperref}

\usepackage{caption, subcaption}
\usepackage{fancyhdr}

\usepackage{siunitx}
\usepackage[square,comma,nonamebreak,compress,numbers]{natbib}

\usepackage{soulpos}
\usepackage{ulem}
\usepackage{authblk} 
\usepackage{float}

\usepackage[inkscapearea=page]{svg}
\usepackage{amsmath}
\usepackage{amssymb}
\usepackage[noabbrev]{cleveref}
\usepackage{upgreek}
\usepackage{adjustbox}
\usepackage{color}
\usepackage{enumitem}
\usepackage{todonotes}

\usepackage{pgfplots}
\usepackage{pgfplotstable}
\usepackage{filecontents}
\usepackage{tikz}
\usepackage{tikzscale}
\usepackage{tikz-3dplot}

\usetikzlibrary{spy}
\usetikzlibrary{intersections}
\usetikzlibrary{arrows,shapes}
\usetikzlibrary{spy}
\usetikzlibrary{backgrounds}  
\usetikzlibrary{decorations}
\usetikzlibrary{decorations.markings}
\usetikzlibrary{positioning}
\usetikzlibrary{patterns}
\usetikzlibrary{calc} 
\usetikzlibrary{quotes} 
\usetikzlibrary{external} 
\usetikzlibrary{matrix}

\pgfplotscreateplotcyclelist{customCycleList}
{%
  {blue, mark=o},
  {red, mark=square},
  {darkgreen, mark=diamond},
  {cyan, mark=diamond},
  {darkbrown, mark=triangle*},
  {black, mark=pentagon},
}

\pgfplotsset{every axis/.append style= {
    cycle list name=customCycleList,
}}

\usepackage{abstract}
\usepackage{placeins}

%% file: 00_frontpage.tex
\title{The Discontinuous Strain Method:\\ accurately representing fatigue and failure}

\author[1,*]{Leon Herrmann}
\author[1,*]{Alireza Danshyar}
\author[1]{Stefan Kollmannsberger}

\affil[1]{Chair of Computational Modeling and Simulation, Technical University of Munich, School of Engineering and Design, Arcisstraße 21, Munich, 80\,333, Germany}

\affil[*]{These authors contributed equally to this work and therefore share the first authorship.}

\newcommand{\publicationDate}{\today}

\date{}
\fancypagestyle{plain}{%
\fancyhf{}%
\vspace{-0.0cm}
\fancyfoot[L]{
\vspace{-0.0cm} 
\footnotesize{Preprint submitted to Computational Mechanics.  \hfill
\publicationDate
\\
}
}
}
